\documentclass[aps,prl,reprint,twocolumn,superscriptaddress,english]{revtex4}
\usepackage{amssymb}
\usepackage{graphicx}
\usepackage{amsmath}
\usepackage{amsfonts}
\usepackage[T1]{fontenc}
\usepackage[latin9]{inputenc}
\usepackage{array}
\usepackage{multirow}
\usepackage{color}
\usepackage{esint}
\usepackage{bm}
\usepackage{color}
\usepackage{bbm}
\usepackage{hyperref}
\usepackage{babel}
\usepackage{titlesec}

\begin{document}


\global\long\def\ud{\mathrm{d}}
\global\long\def\ui{\mathbbm{i}}
\global\long\def\id{\mathbbm{1}}

\title{Dynamical classification of topological quantum phases}

\author{Lin Zhang}
\thanks{These authors contribute equally to this work.}
\affiliation{International Center for Quantum Materials, School of Physics, Peking University, Beijing 100871, China.}
\affiliation{Collaborative Innovation Center of Quantum Matter, Beijing 100871, China.}

\author{Long Zhang}
\thanks{These authors contribute equally to this work.}
\affiliation{International Center for Quantum Materials, School of Physics, Peking University, Beijing 100871, China.}
\affiliation{Collaborative Innovation Center of Quantum Matter, Beijing 100871, China.}

\author{Sen Niu}
\affiliation{International Center for Quantum Materials, School of Physics, Peking University, Beijing 100871, China.}
\affiliation{Collaborative Innovation Center of Quantum Matter, Beijing 100871, China.}

\author{Xiong-Jun Liu
\footnote{Correspondence author: xiongjunliu@pku.edu.cn}}
\affiliation{International Center for Quantum Materials, School of Physics, Peking University, Beijing 100871, China.}
\affiliation{Collaborative Innovation Center of Quantum Matter, Beijing 100871, China.}
\affiliation{Institute for Quantum Science and Engineering and Department of Physics, Southern University of Science and Technology, Shenzhen 518055, China}

\begin{abstract}
Topological phase of matter is now a mainstream of research in condensed matter physics, of which the classification, synthesis, and detection of topological states have brought excitements over the recent decade while remain incomplete with ongoing challenges in both theory and experiment. Here we propose to establish a universal dynamical characterization of the topological quantum phases classified by integers, {and further propose the high-precision dynamical schemes to detect such states.} The framework of {the dynamical classification theory} consists of basic theorems. First, we uncover that classifying a $d$-dimensional ($d$D) gapped topological phase {of generic multibands} can reduce to a ($d-1$)D invariant defined on so-called band inversion surfaces (BISs), rendering a {\it bulk-surface duality} which simplifies the topological characterization. Further, we show in quenching across phase boundary the (pseudo)spin dynamics to exhibit unique topological patterns on BISs, which are attributed to the post-quench bulk topology and manifest a {{\it dynamical bulk-surface correspondence}}. {For this} the topological phase is classified by a dynamical topological invariant measured from dynamical spin-texture field on the BISs. Applications to quenching experiments on feasible models are proposed and studied, {demonstrating the new experimental strategies to detect topological phases with high feasibility.} This work opens a broad new direction to classify and detect topological phases by quantum dynamics.
\end{abstract}

\maketitle

In the last over thirty years, following the discovery of the quantum Hall effect~\cite{Klitzing1980, Tsui1982}, physicists have established a new classification of fundamental states of quantum matter, called topological quantum phases~\cite{Thouless1982,Wen1990}, which are beyond the scope of the celebrated Landau-Ginzburg-Wilson framework that characterizes quantum matter by local symmetry-breaking orders~\cite{Landau1999}. A topological quantum phase can be classified by bulk topological invariant defined in the ground state at equilibrium, and hosts protected boundary modes through the bulk-boundary correspondence~\cite{Hasan2010, Qi2011}.
This characterization crucially influences the strategies of identifying topological states in experiment, which is challenging in general since a non-local topological invariant may not have direct physical measurements.
Despite the great success achieved with various strategies in discovering new topological matter, such as topological insulators~\cite{Konig2007,Hsieh2008,Xia2009,Chang2013} and semimetals~\cite{Xu2015, Lv2015} whose boundary modes can be probed by transport measurement or angle resolved photoemission spectroscopy, by far only a small portion of topological states predicted in theory have been uncovered in experiment~\cite{Benevig2017}. In some circumstances the measurements are not fully unambiguous for being not direct observations of topological numbers, for which there are topological states, e.g., the topological superconductors~\cite{Kitaev2001,Reed-Green2000,Alicea2012,Elliott2014,Sato2017}, even well explored in theory, necessitating further experimental confirmation.

As yet, the topological states have been primarily characterized with classification theories developed for equilibrium systems, ranging from symmetry-protected topological phases~\cite{Chen2012, Chen2013} to intrinsic topological orders~\cite{Wen1990}.
A fundamental question arises that, {\it for a generic topological quantum phase defined {for the equilibrium ground state}, is there a non-equilibrium classification of such phase?} We address this fundamental issue in the present work, and establish a dynamical classification theory for topological quantum phases characterized by integers of {multiband and} all dimensions. Not only being of the clear fundamental significance, the dynamical classification is also experimentally important in probing new topological quantum physics, particularly relevant for ultracold atom systems where, due to heating, the equilibrium ground states are usually hard to achieve, but the quantum dynamics can be readily engineered {as having been considered in the recent studies of topological systems~\cite{quench1,quench2,quench3,quench4,quench5}.} The cold atom experiments have demonstrated the feasibility of simulating novel topological systems with quantum gases, including the 1D Su-Schrieffer-Heeger (SSH) chain~\cite{Su1980,Atala2012}, 1D chiral topological phase~\cite{Liu2013,Song2017}, and 2D Chern insulators~\cite{Jotzu2014, Aidelsburger2015, Wu2016, Sun2017}. {In particular, characterizing the band topology of an equilibrium Hamiltonian from quantum dynamics was proposed and explored very recently in theory and experiment~\cite{Song2017,Tarnowski2017,Wang2017}. Nevertheless, these studies are not universal and valid for two-band models in specific dimensions. No dynamical classification theory was proposed until the present work.}

The framework of the dynamical classification theory is made up of several basic theorems uncovered in this work. First, for a generic $d$-dimensional ($d$D) gapped topological phase characterized by integer invariants, we show a {\it bulk-surface duality} that its classification reduces to a ($d-1$)D invariant defined on the so-called band inversion surfaces (BISs). Further, in quenching from trivial to topological phases, the BISs are captured dynamically that on BISs the time-averaged (pseudo)spin-polarizations vanish in the unitary evolution. Finally, the bulk topology of the post-quench phase is classified by the dynamical topological invariant determined via an emergent dynamical spin-texture field on BISs. The results manifest a dynamical {\it bulk-surface correspondence} which can be directly measured with high feasibility in quenching experiments. {With the dynamical classification theory we propose various high-precision dynamical schemes to detect topological quantum states, of which the application to measuring Chern insulators has been remarkably achieved in a latest experiment~\cite{Sunwei2018}.}

{\it\bf Generic model.}--We start with the generic $d$-dimensional ($d$D) gapped topological phases, including insulators and superconductors, which are classified by integer invariants in the Altland-Zirnbauer (AZ) symmetry classes~\cite{AZ1997,Schnyder2008,Kitaev2009,Chiu2016}. The basic Hamiltonian can be written in the elementary representation matrices of the Clifford algebra~\cite{Morimoto2013, Chiu2013} (see also Supplementary Material~\cite{SI})
\begin{equation}\label{model}
\mathcal{H}(\mathbf{k})=\vec{h}(\mathbf{k})\cdot\vec{\gamma}=\sum_{i=0}^{d}h_{i}(\mathbf{k})\gamma_{i},
\end{equation}
where the $\vec\gamma$ matrices define a (pseudo)spin obeying anti-commutation relation $\left\{ \gamma_{i},\gamma_{j}\right\}=2\delta_{ij}\id$, with $i,j=0,1,\dots,d$, and $\vec{h}(\mathbf{k})$ mimics a $(d+1)$D Zeeman field depending on the Bloch momentum ${\bold k}$ in BZ.
The dimensionality of $\gamma$ matrices in the elementary representation reads $n_d=2^{d/2}$ (or $2^{(d+1)/2}$) if $d$ is even (or odd), which is the minimal requirement to open a topological gap for the $d$D topological phase~\cite{Chiu2013}. In the 1D/2D regimes, for instance, the $\gamma$ matrices simply reduce to the Pauli ones, and $\mathcal{H}(\mathbf{k})$ describes a two-band model for the topological states, e.g., the well-known SSH model~\cite{Su1980, Chiu2016} for 1D BDI class insulator and the Haldane model~\cite{Haldane1988} for 2D Chern insulator. Similarly, for 3D/4D phases, the $\gamma$ matrices take the Dirac forms, and a fully gapped topological phase has to involve at least four bands, such as the 3D AIII class topological insulator (also DIII class supercondoctors)~\cite{Volovik2003,Schnyder2008} and 4D quantum Hall effect~\cite{Zhang2001}. While for convenience the {dynamical classification theory is formulated with the basic Hamiltonian written in the above elementary representation, the theory applies to any generic multiband model of the $d$D phase (see proof in Appendix)}. Therefore, all the main results presented based on Eq.~\eqref{model} directly apply to the generic $d$D gapped topological phases characterized by integer invariants.
\begin{figure}
\includegraphics[width=1.0\linewidth]{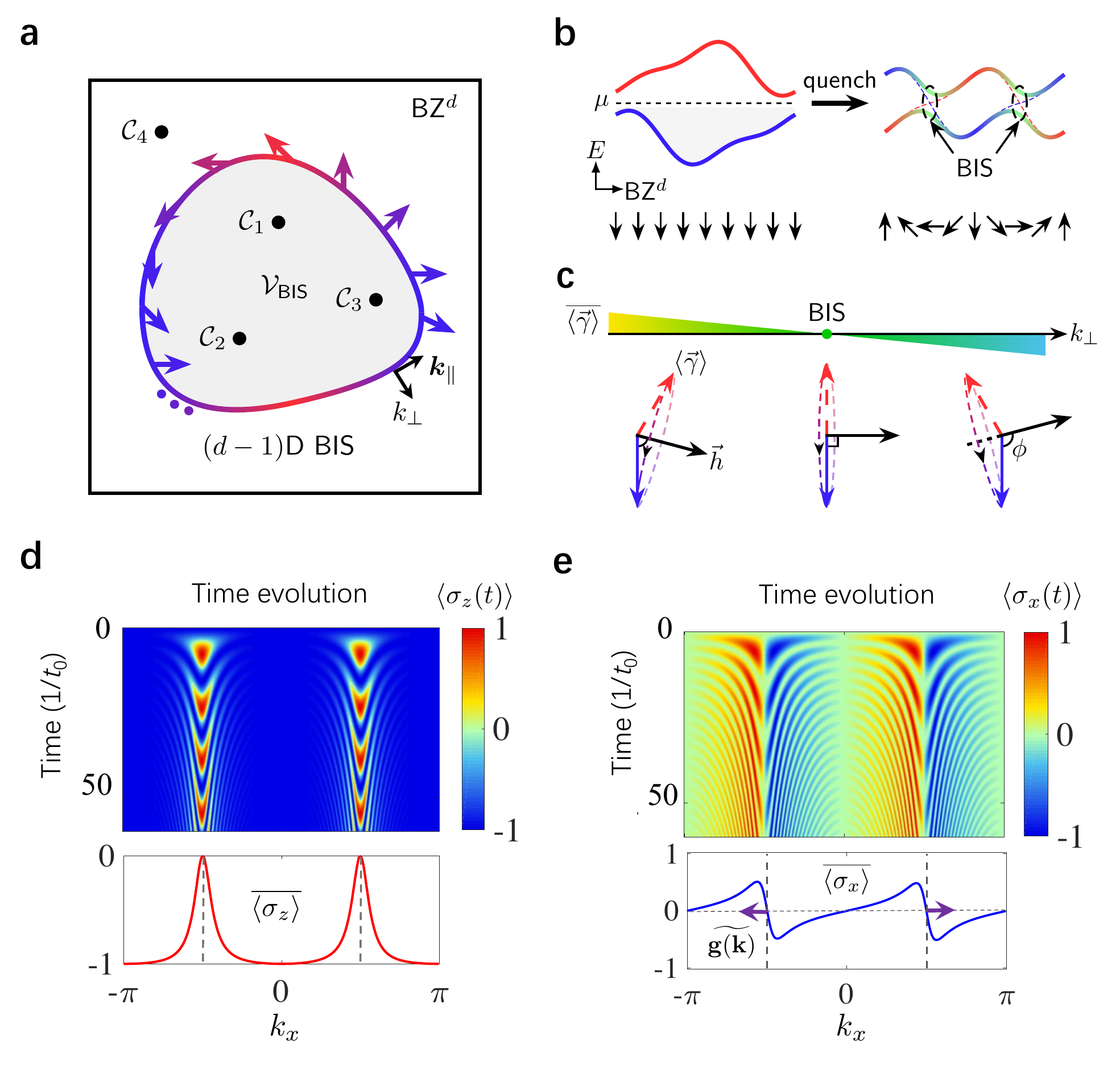}
\caption{{\bf Dynamical bulk-surface correspondence and the 1D exemplification.}
{\bf a}, 
The $d$-dimensional topological phases classified by ${\mathbb Z}$ can be characterized by
the winding of the SO field (colored arrows) on the $(d-1)$-dimensional BISs, which counts the total topological
charges $\mathcal{C}_{i}$ of the SO field in the region $\mathcal{V}_{\mathrm{BIS}}$ (gray) enclosed by BISs with $h_{0}<0$.
Here the momentum ${\bf k}$ is decomposed into $(k_{\perp},\boldsymbol{k}_{\parallel})$
(black arrows) with $k_{\perp}$ perpendicular to the BISs and pointing to the complementary region of $\mathcal{V}_{\mathrm{BIS}}$, i.e., $\bar{\mathcal{V}}_{\mathrm{BIS}}$ with $h_{0}>0$.
{\bf b-c}, Detecting topological patterns on the BISs by quench dynamics.
A quench process is employed,
where a trivial system with highly polarized (pseudo)spins in the negative direction of the axes $h_{0}$ is suddenly quenched to a topological band with
non-trivial spin textures ({\bf b}). Subsequently,
dramatic spin procession occurs near the BISs with the rotation axis being the post-quench vector
field $\vec{h}(\mathbf{k})$ ({\bf c}). The SO field is encoded by the variation of time-averaged
spin polarizations $\overline{\langle\vec{\gamma}\rangle}$ with respect to $k_{\perp}$. Here the blue and red dashed lines in \textsf{\bf b} represent the $h_{0}$ component of the post-quench Hamiltonian with the band crossing being the $(d-1)$D BIS, and $\phi$ in \textsf{\bf c} denotes the relative angle between the ininital spin (blue) and the post-quench vector field $\vec{h}(\mathbf{k})$.
{\bf d-e}, Numerical results for the 1D model. Time evolution
(upper panel) and the corresponding averages (lower
panel) of spin textures are shown after a sudden quench from $m_z=8t_0$ (trivial) to $0$ (topological) with
$t_{\mathrm{so}}=0.2t_{0}$. The BIS is determined by momenta with vanishing time-averaged spin polarizations
$\overline{\langle\sigma_{x,z}\rangle}$ (dashed lines), and the new dynamical spin-texture field $\widetilde{{\bold g}(\mathbf{k})}=-\partial_{k_\perp}\overline{\langle{\sigma_{x}}\rangle}$
(purple arrows in {\bf e}) characterizes the nontrivial topology.}\label{fig1}
\end{figure}

To facilitate the description, we pick up an arbitrary component, say $h_0(\mathbf{k})$, from the vector field $\vec h(\mathbf{k})$ to characterize the `dispersion' of the $n_d$ decoupled bands. Accordingly, we denote the remaining components of $\vec h(\mathbf{k})$ as the `spin-orbit' (SO) field ${\boldsymbol{h}}_{\rm so}(\bold k)=(h_1,...,h_d)$. Without the SO field, the band crossing occurs for the $n_d$ decoupled bands if $h_0(\bold k)=0$ results at certain momentum points of the BZ. All such momentum points form closed $(d-1)$D surface, dubbed {\it band inversion surface} (BIS), which is a key {concept} in this work. {The BIS is related to, but conceptually different from the
familiar phenomenon of band inversion in topological insulators, and can be defined for broad classes of phases including topological superconductors.} Across the BIS the energy difference between half of the $n_d$ bands and the remaining half switches sign. In general one can have multiple BISs in the first BZ. The gap opens when the nonzero SO field ${\boldsymbol{h}}_{\rm so}(\bold k)$ is switched on in the BISs. As shown below, the $d$D gapped phase being topologically nontrivial necessitates that the BZ includes at least one BIS which hosts a nonzero $(d-1)$D invariant.

{\bf Bulk-surface duality.}--The topology of the Hamiltonian ${\cal H}(\bold k)$ is classified by $d$D winding number for odd dimensionality $d=2n-1$, or the $n$-th Chern number if $d=2n$. The topological characterization of the equilibrium phase is formulated by a mapping from the BZ, i.e. a $d$D torus $T^d$, to the $d$D spherical surface $S^d$ through the unit vector field $\bold n(\bold k)=\vec h/|\vec h|$. The topological number counts the times that the mapping covers $S^d$. {Interestingly}, as detailed in the Appendix, we show that the topological characterization by {$d$D winding number (or Chern number)} reduces to a {$(d-1)$D Chern number (or winding number)} $w_{d-1}$ defined on the BISs
\begin{equation}\label{theorem1}
{\nu_d(h_0,\bold h_{\rm so})=w_{d-1}(\bold h_{\rm so}),}
\end{equation}
{where $w_{d-1}$ characterizes the integer times that the unit SO field $\hat{\boldsymbol{h}}_{\rm so}(\bold k)=\boldsymbol{h}_{\rm so}/|\boldsymbol{h}_{\rm so}|$ covers over the ($d-1$)D spherical surface $S^{d-1}$ when $\bold k$ runs over all the BISs of the BZ.} Without giving details, we show the above invariant $w_{d-1}$ intuitively by clarifying the key ingredients of the theorem, while the rigorous proof can be found in the Appendix. As for topological phase, the mapping $T^d\mapsto S^d$ covers nonzero integer times of the $d$D spherical surface, any component of the vector field $\vec h(\bold k)$, including $h_0(\bold k)$, spans both positive and negative values in the BZ. Thus the BIS with $h_0(\bold k)=0$ must be obtained. It is convenient to denote by $\bold{\cal V}_{\rm BIS}$ the vector volume of the BZ with $h_0(\bold k)<0$, and $\bar{\bold{\cal V}}_{\rm BIS}$ the volume with $h_0(\bold k)>0$. Note that $w_{d-1}$ in the right hand side of Eq.~\eqref{theorem1} is given by a $(d-1)$D surface integral enclosing the {open} vector volume $\bold{\cal V}_{\rm BIS}$ {(Appendix)}, {which is gauge-independent} and counts the total topological charges located at the singularities with $\bold h_{\rm so}(\bold k)=0$ in this volume (see Fig.~1a). This result can be intuitively understood in the following way, for which we rewrite that $h_0(\bold k)=m_0+{\tilde h}_0(\bold k)$, with ${\tilde h}_0$ being momentum-dependent and $m_0$ mimicking a constant `magnetization' which tunes the topology of the phase. It is ready to know that when $m_0>\max[|{\tilde h}_0(\bold k)|]$ for all $\bold k$, the phase is trivial since no BIS can be obtained in the BZ. The BIS starts to emerge in the BZ by reducing the magnetization to $m_0\lesssim\max[|{\tilde h}_0(\bold k)|$, while the gap of the system keeps open if no topological charges (at $\bold h_{\rm so}(\bold k)=0$) enter the vector volume $\bold{\cal V}_{\rm BIS}$. Further reducing $m_0$ eventually enables that the topological charges pass through the BIS and enter the volume $\bold{\cal V}_{\rm BIS}$. Note that when a topological charge crosses the BIS, the bulk gap closes one time since $h_0=\bold h_{\rm so}=0$ occurs at the crossing momentum on the BIS. This leads to a topological phase transition with the topological invariant varying by one. Then the number of topological charges with $\bold h_{\rm so}(\bold k)=0$ enclosed in the volume $\bold{\cal V}_{\rm BIS}$ reflects the times of gap closing by tuning $m_0$ from trivial to the final topological regime, and the total topological charge gives the bulk topological number of the phase, as characterized by the theorem in Eq.~\eqref{theorem1}.

The above theorem shows a bulk-surface duality which maps the classification of bulk topology to the characterization on BISs. The bulk-surface duality provides a simplified characterization of the topological phases with lower-dimensional invariants, {while there are two important issues having to be addressed. First, the BISs are not unique but depends on the choice of $h_0$ axis. Then how to precisely identify BISs in a real experiment? Further, how to measure the topology defined on the BISs?} As we show below that, the uncovered bulk-surface duality is further mapped to {a novel dynamical form} in the quench dynamics, {which directly measures all the topological features on BISs with high feasibility.}

{\it\bf Non-equilibrium classification: {dynamical bulk-surface correspondence.}}--We proceed to consider the non-equilibrium characterization of topological phases based on the bulk-surface duality shown above. The quantum dynamics can be induced by quenching the phase from a deep initially trivial phase at time $t=0$, which is given by setting $m_0|_{t<0}\gg\max[|\vec h(\bold k)|]$, to the final topological state which is obtained in proper regime $|m_0|\bigr|_{t>0}<\max[|h_0(\bold k)|]$ (Fig.~1b). The unitary evolution of the (pseudo)spin $\vec \gamma$ is then governed by the post-quench Hamiltonian ${\cal H}(m_0,\bold k)$. The quenching dynamics can be quantified by introducing the dynamical averaging of the spin-polarization, which is obtained by
\begin{equation}\label{dynamical1}
\overline{\left\langle \gamma_{i}(\bold k)\right\rangle }=\lim_{T\to\infty}\frac{1}{T}\int_{0}^{T}\ud t\,\mathrm{Tr}\left[\rho_{0} e^{\ui\mathcal{H}t}\gamma_{i}e^{-\ui\mathcal{H}t}\right],
\end{equation}
where $\rho_0$ is density of matrix of the initial state. The peculiar effects are obtained on the BISs. Note that the system is initialized in the fully spin-polarized phase with the (pseudo)spin of all Bloch states pointing along negative $h_0$ axis (for positive $m_0$). The spin processes with respect to $\vec h(\bold k)$ after quenching. On the BISs, the vector $\vec h(\bold k)=\bold h_{\rm so}(\bold k)$ is perpendicular to initial spin polarization, which leads to the procession of the spin vector $\vec\gamma$ within the plane perpendicular to $\bold h_{\rm so}$ while incorporating $h_0(\bold k)$ axis (Fig.~1c). As a result, the dynamical averaging $\overline{\left\langle \gamma_{i}\right\rangle }$ vanishes right on the BISs for all $i$-components. On the other hand, at the momentum $\bold k$ away from the BISs the $h_0$ component is nonzero, so $\vec h$ is not perpendicular to the initial spin polarization. Then the spin procession leads to a nonzero $\overline{\left\langle \vec\gamma\right\rangle }$ along the $\vec h(\bold k)$-direction (Fig.~1c). With these results we conclude a dynamical characterization of the BISs that
\begin{equation}\label{BIS}
\overline{\left\langle \gamma_{i}(\bold k)\right\rangle }=0, \ \ \mbox{for} \ \bold k\in\mbox{BISs}, \ i=0,1,2,...,d.
\end{equation}
The above characterization can also be understood that on BISs, the resonant spin-reversing transitions $\vec\gamma\rightarrow-\vec\gamma$ are induced by the SO field $\bold h_{\rm so}(\bold k)$ due the vanishing gap (for band-crossing surfaces). Thus the time-averaged spin-polarizations vanish at BISs. In contrast, away from the BISs, the nonzero $h_0(\bold k)$ severs as a detuning for the spin-reversing transitions, so the time-averaged spin-polarizations do not vanish.

The vanishing $\overline{\left\langle \gamma_{i}\right\rangle }$ on BIS implies that the topological number of the $d$D gapped phase cannot be obtained via the time-averaged spin-texture $\overline{\left\langle \vec\gamma(\bold k)\right\rangle}$. Interestingly, more nontrivial features of the quench dynamics are captured by the {\it variation} of the time-averaged spin polarizations across the BISs. Note that the component $h_0<0$ in the vector volume ${\cal V}_{\rm BIS}$ and $h_0>0$ in the volume $\bar{\cal V}_{\rm BIS}$. When passing through a BIS, the relative angle $\phi$ between the initial spin-polarization and the vector $\vec h(\bold k)$ varies from $\phi<\pi/2$ in ${\cal V}_{\rm BIS}$, with a nonzero dynamical averaging $\overline{\left\langle \vec\gamma\right\rangle }$ pointing to $\vec h(\bold k)$-direction, to $\phi>\pi/2$ in $\bar{\cal V}_{\rm BIS}$, with nonzero $\overline{\left\langle \vec\gamma\right\rangle }$ pointing oppositely to $\vec h(\bold k)$-direction (Fig.~1c). This follows that the variation of $\overline{\left\langle \vec\gamma\right\rangle}$ across the BIS follows the direction of the spin-orbit field $\bold h_{\rm so}$. To quantify this picture, we define a new dynamical spin-texture field $\widetilde{\bold g(\bold k)}$, whose components are given by $\widetilde{g_i(\bold k)}\equiv-\frac{1}{{\cal N}_{\bold k}}\partial_{k_\perp}\overline{\left\langle \gamma_i\right\rangle}$, with ${\cal N}_{\bold k}$ being the normalization factor. Here $k_\perp$ denotes the momentum perpendicular to BIS and points from ${\cal V}_{\rm BIS}$ to $\bar{\cal V}_{\rm BIS}$. It can be shown from Eq.~\eqref{dynamical1} that on the BISs
\begin{eqnarray}\label{dynamical2}
\widetilde{g_i(\bold k)}\bigr|_{\bold k\in\mbox{BISs}}=\hat h_{{\rm so},i}(\bold k).
\end{eqnarray}
With this identity we reach immediately that the topological invariant defined on the BISs can be recast into a dynamical form, namely
\begin{equation}\label{dynamical3}
w_{d-1}=\sum_{j}\frac{\Gamma(d/2)}{2\pi^{d/2}}
\frac{1}{\left(d-1\right)!}\int_{\mathrm{BIS}_j}\widetilde{\bold g(\bold k)}\bigr[\ud\widetilde{\bold g(\bold k)}\bigr]^{d-1}.
\end{equation}
The right hand side of the above formula describes the mapping from BISs to $(d-1)$D spherical surface which the emergent dynamical field $\widetilde{\bold g(\bold k)}$ belongs to, rendering the dynamical classification of the $d$D gapped phase. The dynamical topological invariant is intuitively described by the coverage of the dynamical spin-texture $\widetilde{\bold g(\bold k)}$ over the full $(d-1)$D spherical surface.
The results in Eqs.~\eqref{BIS}-\eqref{dynamical3} manifest a highly nontrivial {\it dynamical bulk-surface correspondence}, which can be directly measured from the dynamical spin-polarization patterns emerging on the BISs. {Since the quantum spin dynamics is resonant and nontrivial only on BISs, the dynamical bulk-surface correspondence can be well resolved in experiments. This enables us to propose high-precision dynamical schemes to detect the bulk topology, as studied below with experimentally feasible models.

We note that the dynamical classification theory is essentially based on the mapping of the $d$D bulk topology to $(d-1)$D BISs. This feature is valid for generic multiband systems, in which case the BISs are defined among every $n_d$ bands in the BZ. The topology only relies on the $(d-1)$D invariants `locally' defined on the BISs, which can be measured based on a sequence of quenches by initializing the system in different bands (Appendix). This dimension reduction is the essential difference from the previous dynamical schemes which are applicable to two-band models~\cite{Song2017,Tarnowski2017,Wang2017}.}

{\bf Application to quenching experiments.}--We first consider the 1D topological phases of AIII class obtained by the Hamiltonian $\mathcal{H}(k_x)=\vec{h}(k_x)\cdot\vec{\sigma}$, where $h_x=t_{\mathrm{so}}\sin k_x\equiv h_{\rm so}, h_y=0$, and $h_z=m_{z}-t_{0}\cos k_x\equiv h_0$, with $t_{0}$ and $t_{\rm so}$ denoting the nearest-neighbor spin-conserved and spin-flipped hopping coefficients. The model was proposed in Ref.~\cite{Liu2013} and realized recently in experiment for $^{173}$Yb fermions~\cite{Song2017}. The effective magnetization $m_{z}$ can be precisely tuned by a bias magnetic field. The quench study is performed by tuning the Zeeman term from
$m_{z}\gg t_{0}$, which corresponds to the deep trivial
regime with the system initialized in spin-down, to $\left|m_z\right|<t_{0}$, which lies in topological regime.
The quantum dynamics is then governed by the unitary evolution operator (see Methods).
The numerical results are shown in Fig.~1d-e, with the parameters taken as $t_{\mathrm{so}}=0.2t_{0}$. The resonant spin-flip transitions driven by SO field emerge at two momentum points $k_{\rm BIS}=\pm\cos^{-1}(m_z/t_0)$ with vanishing time-averaged spin polarizations $\overline{\langle\vec\sigma\rangle}$, which manifests the BIS (band inversion points). Away from the BIS the spin-flip oscillation is well suppressed by the energy gap. The opposite magnitudes of the directional derivative $\partial_{{k}_{\perp}}\overline{\left\langle \sigma_{x}\right\rangle }$ at two band inversion points show the nontrivial topology of the post-quench regime, and gives the winding number $+1$ for $|m_z|<t_0$ (Fig.~1e).

\begin{figure}
\includegraphics[width=1.0\linewidth]{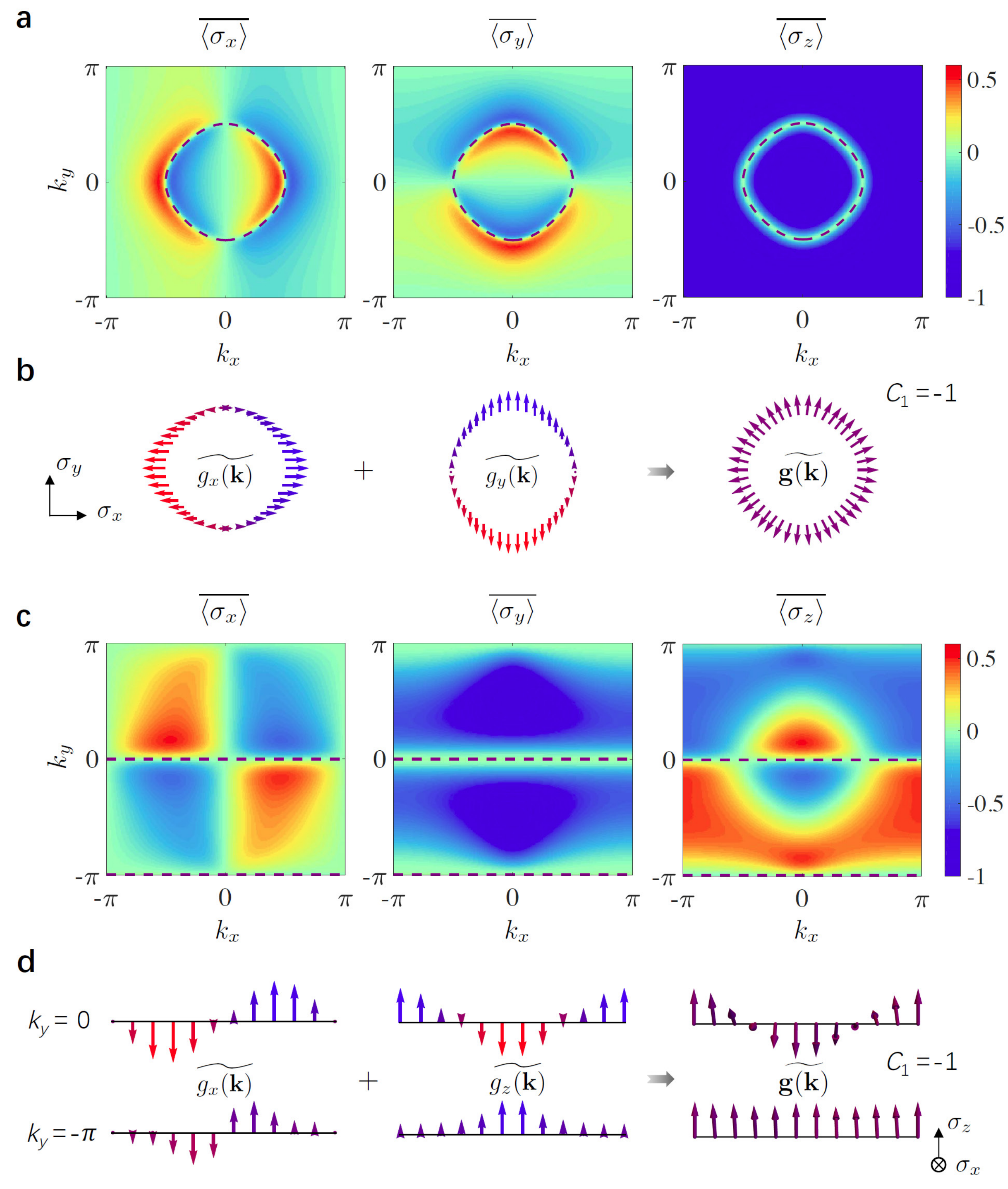}
\caption{{\bf Characterizing 2D Chern insulators in different quench processes}.
{\bf a-b}, Topological patterns by quenching along $h_z({\bf k})$ axis.
Time-averaged spin textures ({\bf a}) are measured after a sudden quench from $m_z=8t_0$ to $t_0$ ($m_x=m_y=0$, $t_{\rm so}^{x,y}=0.2t_{0}$),
with BIS being characterized by $\overline{\langle\vec\sigma\rangle}=0$ (green ring).
The dynamical field $\widetilde{{\bold g}(\mathbf{k})}$ obtained from {\bf a} indicates
the Chern number $C_1=-1$ ({\sf\textbf{b}}).
\textsf{\textbf{c-d}}, Topological patterns by quenching $h_y({\bf k})$ axis. Time-averaged spin textures ({\bf c})
and the dynamical field on the BIS ({\sf\textbf{d}}) are shown after a quench from $m_y=50 t_0$ to $0$, with $m_z=t_0, m_x=0$, and $t_{\rm so}^x=0.5t_{\rm so}^y=t_0$.
Here the two BISs are at $k_y=0,-\pi$. While the winding of $\widetilde{{\bold g}(\mathbf{k})}$ is trivial along $k_y=-\pi$, the non-zero
winding number along $k_y=0$ characterizes the topological phase with the Chern number $C_1=-1$.}\label{fig2}
\end{figure}

In 2D case we consider the quantum anomalous Hall model~\cite{Liu2014,Wang2018} realized in recent experiments~\cite{Wu2016,Sun2017} $\mathcal{H}(\mathbf{k})=\vec{h}(\mathbf{k})\cdot\vec{\sigma}$,
where the vector field reads $\vec{h}(\mathbf{k})=(m_x+t_{\mathrm{so}}^{x}\sin k_{x},m_y+
t_{\mathrm{so}}^{y}\sin k_{y},m_{z}-t_{0}\cos k_{x}-t_{0}\cos k_{y})$.
The flexibility in the decomposition of the vector $\vec{h}(\mathbf{k})$ allows us to consider different quench ways, e.g., by quenching $h_{z}$ or $h_{y}$ axis, {which can benefit real experimental measurement.}
First, we take $h_0\equiv h_z, \bold h_{\rm so}\equiv(h_x,h_y)$, and the quench is performed by varying $m_z$ suddenly from $m_z\gg t_0$ to $0<|m_z|<2t_0$, while setting $m_x=m_y=0$. Similar to 1D model, the resonant transition between the spin-down and spin-up states only occurs at the BIS (band inversion ring). From the numerical results given in Fig.~2a, with $m_z=t_0$ and $t_{\rm so}^{x,y}=0.2t_0$, the vanishing time-averaged spin-polarizations $\overline{\langle\sigma_i\rangle}=0$ are obtained on the BIS. Further, the emergent dynamical field $\widetilde{\bold g(\bold k)}=-\partial_{{k}_\perp}\overline{\langle{\vec\sigma}\rangle}$ exhibits a nonzero 1D winding on the BIS, which characterizes a topological charge located at $\Gamma$ point [with $\bold k=(0,0)$] and gives the Chern number $C_1=-1$ (Fig.~2b). Similarly, if $-2t_0<m_0<0$ for the post-quench regime, an opposite dynamical winding exhibits on the BIS which circles $M$ point [with $\bold k=(\pi,\pi)$], and gives $C_1=+1$ (see Appendix). Secondly, we take $h_0\equiv h_y$ and quench the $h_{y}$ axis by ramping $m_y$ quickly from $m_y\gg t_0$ to $m_y=0$, while setting $m_x=0$ and $m_z=t_0$. The time-averaged dynamical spin textures exhibit line-shape structures (Fig.~2c), which correspond to the two open BISs at $k_y=0$ and $-\pi$, respectively. Despite a trivial pattern along $k_x$ with $k_{y}=-\pi$, the winding of the dynamical spin-texture $\widetilde{\bold g(\bold k)}$ (in $x-z$ plane) along $k_x$ with $k_{y}=0$ measures the Chern number $C_1=-1$ (Fig.~2d). Applying the dynamical classification to measuring topological states with high Chern numbers is straightforward and is found in Appendix.

\begin{figure}
\includegraphics[width=1.0\linewidth]{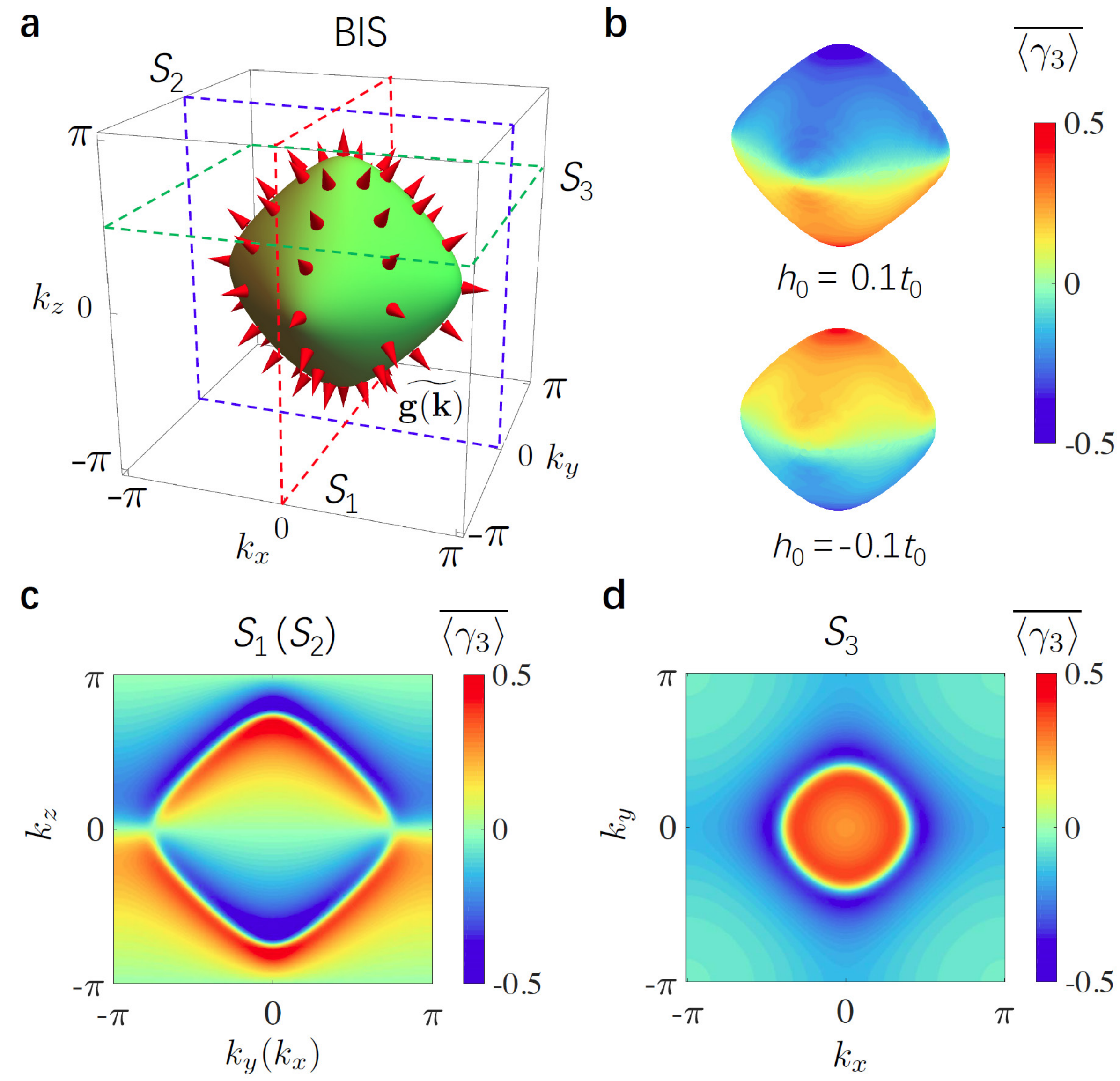}
\caption{{\bf 3D chiral topological phases}.
{\bf a}, The BIS (green sphere) defined by $h_0({\bf k})=0$ and {the dynamical (pseudo)spin-texture field $\widetilde{{\bold g}(\mathbf{k})}$ (red arrows)
are shown for $m_z=1.3t_0$ and $t_{\rm so}=0.2t_0$. {\bf b}, In the practical measurement the $\widetilde{{\bold g}(\mathbf{k})}$ field can be determined through the variation of spin textures at two equal-energy surfaces close to the BIS, e.g. at $h_0=\pm0.1t_0$.}
{\bf c-d}, Time-averaged spin textures $\overline{\langle\gamma_3\rangle}$ in cross sections $k_{x,y}=0$ ($S_{1,2}$ in {\bf c})
and $k_z=\frac{\pi}{2}$ ($S_3$  in {\sf\textbf{d}}). The quench is taken from $m_z=8t_0$ to $1.3t_0$
and $\overline{\langle\gamma_{1,2}\rangle}$ are shown in Appendix.}\label{fig3}
\end{figure}

We further consider the application to 3D topological phases, whose hosting Hamiltonian reads $\mathcal{H}(\mathbf{k})=\vec{h}(\mathbf{k})\cdot\vec{\gamma}$,
with $h_{0}(\mathbf{k})=m_{z}-t_{0}\sum_{i}\cos k_{i}$ and
$h_{i}=t_{\mathrm{so}}\sin k_{i}$ ($i=x,y,z$). Here we take that
$\gamma_{0}=\sigma_{z}\otimes\tau_{x}$, $\gamma_{1}=\sigma_{x}\otimes\id$,
$\gamma_{2}=\sigma_{y}\otimes\id$ and $\gamma_{3}=\sigma_{z}\otimes\tau_{z}$, where the Pauli matrices
$\sigma_i$ and $\tau_i$ may refer to the real- and pseudo-spin degrees of freedom (e.g., sublattices or orbitals). The Hamiltonian has a chiral symmetry defined by $\sigma_z\otimes\tau_y$, so it belongs to AIII class and is classified by 3D winding numbers in equilibrium theory~\cite{AZ1997,Schnyder2008,Kitaev2009}. The trivial phase corresponds
to $\left|m_{z}\right|>3t_{0}$, while the topological phases include three regions:
(I) $t_{0}<m_{z}<3t_{0}$ with winding number $\nu_3=-1$; (II) $-t_{0}<m_{z}<t_{0}$
with $\nu_3=2$; and (III) $-3t_{0}<m_{z}<-t_{0}$ with $\nu_3=-1$.
We consider the quench from $m_z\gg t_0$ to say phase I ($m_z=1.3t_0$ and $t_{\mathrm{so}}=0.2t_{0}$), with the numerical results being partially presented in Fig.~3 (more is shown in Appendix).
The BIS of the post-quench band is given in Fig.~3a, {with the vector arrows denoting the dynamical field $\widetilde{\bold g(\bold k)}$. In a practical measurement, by resolving the time-averaged spin polarizations $\overline{\langle\vec\gamma\rangle}$ on two closed surfaces sightly inside (with $h_0=-0.1t_0$) and outside (with $h_0=0.1t_0$) the BIS, one can qualitatively determine the vector field $\widetilde{\bold g(\bold k)}$ (without affecting measurement of topology)} from the subtraction of $\overline{\langle\vec\gamma\rangle}$ between the two surfaces. A few examples for $\overline{\langle\gamma_3\rangle}$ are given in Fig.~3b-d, similar for $\overline{\langle\gamma_{1,2}\rangle}$. {From the dynamical field $\widetilde{\bold g(\bold k)}$ on the BIS, as illustrated with arrows in Fig.~3a, which gives a nonzero Chern number $C_1=-1$ on the 2D BIS, corresponding to the 3D winding number $\nu_3=-1$ of the post-quench 3D system.}

\begin{figure}
\includegraphics[width=1.0\linewidth]{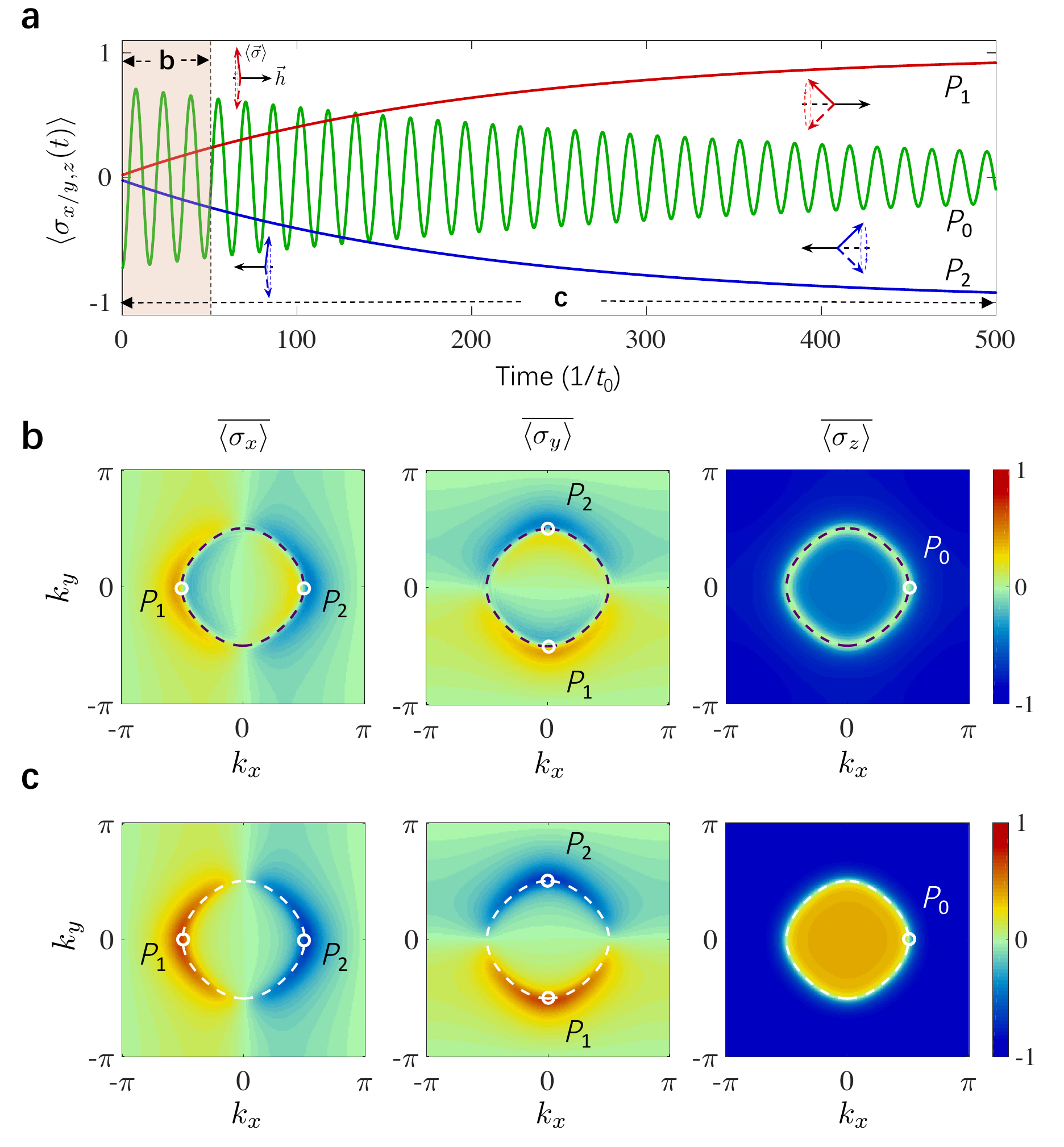}
\caption{{\bf Classifying topology via dissipative dynamics}.
{\bf a}, Post-quench dissipative dynamics of spin polarizations $\langle\sigma_{x,y}(t)\rangle$ (red and blue lines)
for points $P_{1,2}$ and $\langle\sigma_{z}(t)\rangle$ (green line) for $P_{0}$ at finite temperature. Due to dissipation,
the spin $\langle\vec{\sigma}\rangle$ gradually decays to the opposite direction of the procession axis $\vec{h}(\bold k)$ (see insets).
The BIS and topological patterns can be determined by time-averaged spin textures over a short period (see {\bf b})
and a long period (see {\bf c}), respectively.
{\bf b}, Time-averaged spin textures $\overline{\langle\sigma_{x,y,z}\rangle}$ over a period of $50/t_0$,
which resemble the results in Fig.~2a. The BIS is determined by $\overline{\langle\vec{\sigma}\rangle}\simeq0$ (black dashed line).
{\bf c}, Time-averaged spin textures $\overline{\langle\sigma_{x,y,z}\rangle}$ on the BIS over a long term of $500/t_0$ directly reflect the SO field and thus the topological number.
Here we take the temperature $k_{\rm B}T=10t_0$ and the decay rate $\eta=0.005$
with other parameters the same as in Fig.~2a.}\label{fig4}
\end{figure}
While the above study is based on unitary quench dynamics, the dynamical classification can be generalized to the dissipative quantum dynamics at finite temperature. In the presence of dissipation, we expect that quantum dynamics should be nearly unitary within a short-time evolution after quench, while the dissipation effects shall dominate in the long-term evolution. Taking the 2D model as an example,
we describe the dynamics by the Lindblad master equation~\cite{Lindblad,quench3}
\begin{equation}
\frac{\ud}{\ud t}\rho_{\mathbf{k}}=-\ui\left[\mathcal{H},\rho_{\mathbf{k}}\right]+\eta\left(\tilde{\sigma}_{-}\rho_{\mathbf{k}}\tilde{\sigma}_{+}
-\frac{1}{2}\left\{ \tilde{\sigma}_{+}\tilde{\sigma}_{-},\rho_{\mathbf{k}}\right\} \right)
\end{equation}
with initial state being $\rho_{\mathbf{k}}(0)=f(E_{+},T)\left|+,\mathbf{k}\right\rangle \left\langle +,\mathbf{k}\right|
+f(E_{-},T)\left|-,\mathbf{k}\right\rangle \left\langle -,\mathbf{k}\right|$.
Here the distribution function $f(E,T)=\left[e^{\left(E-\mu\right)/k_{\mathrm{B}}T}\pm1\right]^{-1}$ if the system is simulated with fermions (for $+$) and bosons (for $-$), $\left|\pm,\mathbf{k}\right\rangle $
are the eigenstates of the pre-quench Hamiltonian,
the Pauli matrices $\tilde{\sigma}_{\pm}\equiv(\tilde{\sigma}_{x}\pm\ui\tilde{\sigma}_{y})/2$
are defined in the eigenbasis of the post-quench Hamiltonian
$\mathcal{H}$, and $\eta$ is the decay
rate. A numerical study with typical parameters is shown in Fig.~4. For a short-term evolution, we find that the dissipative effect is negligible  (nearly unitary), as demonstrated in Fig.~4a (gray area) and b. In this regime, the previous dynamical classification is well applicable to measure the locations of BISs and the bulk topology (Fig.~4b).
Further, after a long-term evolution, the dissipation
drives the occupation of Bloch states to approach equilibrium results. In this case we find that the spin-polarization $\overline{\langle\vec\sigma(\bold k)\rangle}$ gradually points oppositely to the $\bold h_{\rm so}(\bold k)$-direction on BISs of the post-quench system (Fig.~4a). Then the Chern number can be directly read out by the long-term spin texture $\overline{\langle\vec\sigma(\bold k)\rangle}$, without the further adoption of $\widetilde{\bold g(\bold k)}$ for the classification (Fig.~4a,c).

{\bf Discussion and outlook.}--The dynamical classification exhibits essential advantages in experimental investigation of topological quantum states. In particular, applying the classification theory proposed in this work to measuring 2D Chern insulators has been achieved in a very recent experiment~\cite{Sunwei2018}, which confirms that observing topological phases based on the current dynamical classification is of much higher precision, compared with that based on equilibrium classifications. The advantages are rooted in two essential aspects. First, the bulk-surface duality uncovered here maps the classification of bulk topology to lower-dimensional invariants on BISs, and simplifies the topological characterization. {Secondly, the spin dynamics is resonant and nontrivial only on the BISs, so the dynamical bulk-surface correspondence can be easily resolved in the quench study. Finally, the dynamical scheme by nature is robust against the non-ideal conditions in both the state initialization and measurement. In particular, the system starts at deep trivial regime, with the initial phase being dependent on only a single parameter, Zeeman term, immune to non-ideal conditions. Quenching it to topological regime induces quantum dynamics which exhibits resonant oscillations on the BISs. The short-term unitary quench dynamics characterizing the topology of the post-quench system is not affected by detrimental effects like the thermal effects}, leading to high-precision measurement of full topological phase diagrams, as confirmed in experiment~\cite{Sunwei2018}. In comparison, measuring the topology of static phases necessitates the systems to be carefully initialized, with the quantum states being properly occupied. The preparation is intrinsically sensitive to non-ideal conditions including thermal effects, especially unsatisfactory in the regime close to phase boundaries~\cite{Sun2017}. Our dynamical classification provides new approaches with high feasibility in exploring topological quantum physics.

{Moreover, the dynamical classification does not require the initial trivial phase to be fully polarized along certain direction. In the generic case with a trivial initial phase which has trivial but momentum-dependent (pseudo)spin texture, one can rotate the initial (pseudo)spin of each Bloch state to the direction of $h_0$ axis by a momentum-dependent unitary transformation. Accordingly, under the transformation the new (pseudo)spin bases become momentum-dependent but still topologically trivial. Then, if expressed in the new (pseudo)spin bases, all the results including the characterization of BISs by vanishing time-averaged spin polarization and dynamical bulk-surface correspondence take the same forms as in the case with a fully polarized initial phase. However, taking the fully polarized (deep trivial) phase as the initial state is most convenient for real experiments.}

Finally, this work provides important insights into developing {non-equilibrium} classification for the complete classes of {equilibrium} topological quantum states, as conceptually different from the known various equilibrium classification theories. The dynamical classification theory established here covers different categories of the topological phases characterized by integer invariants, including the A, AIII, BDI, C, CII, and D classes in the AZ ten-fold ways~\cite{AZ1997}. It is promising and of great interests to generalize the current classification theory to all ten-fold categories~\cite{AZ1997,Schnyder2008,Kitaev2009}, crystalline topological states, including the insulators and superconductors protected by space groups~\cite{Slager2013,Chiu2016}, and further to the correlated topological states of fermions and bosons~\cite{Chen2012,Chen2013}. {We note that for the correlated topological phases whose topology can be characterized by quasiparticles or mean-field picture, the concept of BISs can still be well defined and the present dynamical classification theory should be applicable. In turn, the correlation effects can have non-trivial effects on the dynamical bulk-surface correspondence and deserve in-depth studies.}. We therefore expect that the present work can open a broad new direction to classify and detect topological quantum states by non-equilibrium quantum dynamics.
\bigskip

This work was supported by the National Key R\&D Program of China (2016YFA0301604), National Nature Science Foundation of China (under grants No. 11574008 and No. 11761161003), and the Thousand-Young-Talent Program of China.




\section{Appendix A: Characterizing $d$D Topological Phases by $(d-1)$D Invariants}\label{sec1}
\renewcommand{\theequation}{A\arabic{equation}}
\setcounter{equation}{0}  

In this section, we present the rigorous proof of the bulk-surface duality that $d$D topological phases can be characterized by
the $(d-1)$D topological invariant defined on the BISs, including the generalization to multiband systems.\\

\subsection{1. Bulk-surface duality: the $(d-1)$D topological invariant on the BISs}

\begin{figure*}
\includegraphics{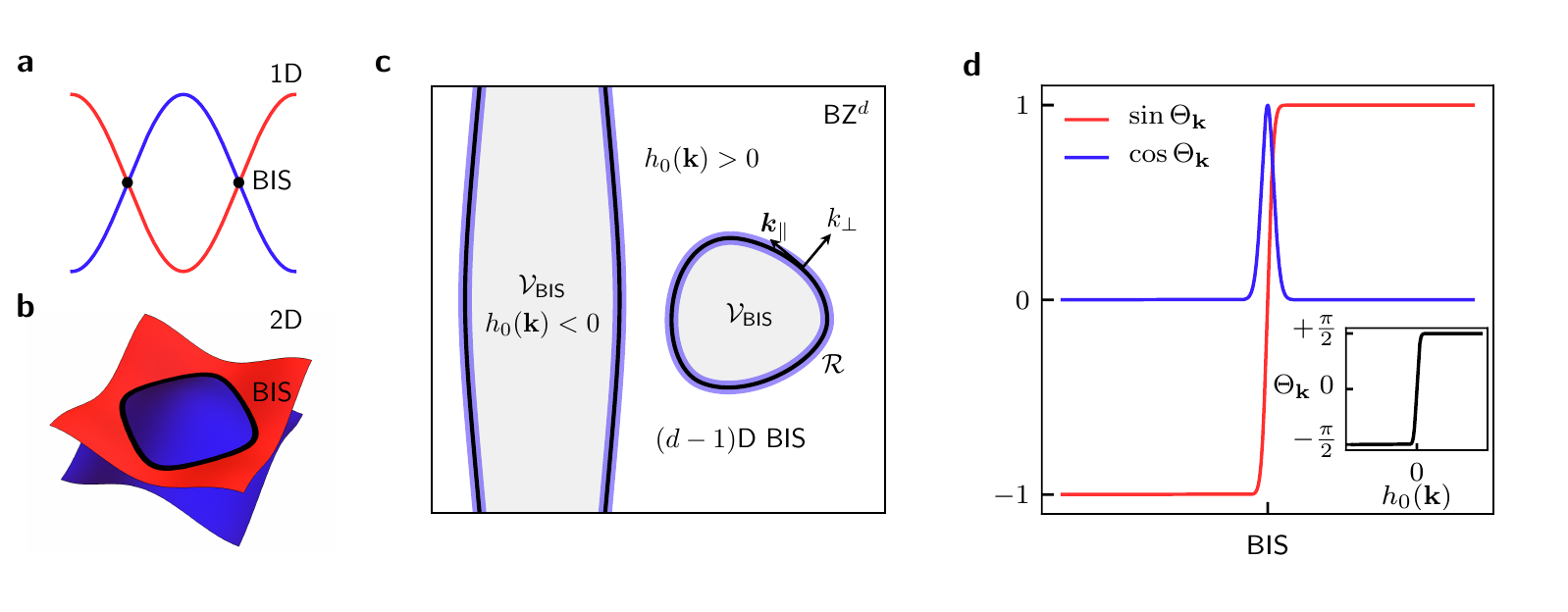}
\caption{\textsf{\textbf{The schematic diagram of BISs and the deformation technique}}. \textsf{\textbf{a-b}}, BISs in 1D and 2D. In 1D, the BIS (black dots) composes of discrete points (\textsf{\textbf{a}}), and becomes a 1D closed curve (the black ring) in 2D (\textsf{\textbf{b}}). \textsf{\textbf{c}}, $(d-1)$D BISs in $d$D BZ. Here $\mathcal{V}_{\mathrm{BIS}}$ denotes the region enclosed by BIS with $h_{0}(\mathbf{k})<0$ (gray). The momentum $\mathbf{k}$ is decomposed into $(k_{\perp},\boldsymbol{k}_{\parallel})$ with $k_{\perp}$ perpendicular to the contour of $h_{0}(\mathbf{k})$ and pointing to the side $\bar{\mathcal{V}}_{\mathrm{BIS}}$ with $h_{0}(\mathbf{k})>0$. \textsf{\textbf{d}}, The function $\sin\Theta_{\mathbf{k}}$ (red) and $\cos\Theta_{\mathbf{k}}$ (blue) used in the deformed Hamiltonian (\ref{eq:s_deformedH}). The insert shows the function $\Theta_{\mathbf{k}}$. The region with non-zero SO field ($\cos\Theta_{\mathbf{k}}$) is denoted by $\mathcal{R}$ (blue) in \textsf{\textbf{c}}.
\label{fig:fig5} }
\end{figure*}

For a generic $d$D gapped topological phases described by the basic Hamiltonian~(\ref{model}),
there can be multiple BISs defined by $\mathrm{BIS}\equiv\{\mathbf{k}|h_{0}(\bold k)=0\}$ in the BZ, being closed to connected BZ boundary (Fig.~\ref{fig:fig5}a-c). For convenience, here we denote
by $\mathcal{V}_{{\rm BIS}}$ the region enclosed by BISs with $h_{0}(\mathbf{k})<0$,
and decompose the momentum $\mathbf{k}$ into $(k_{\perp},\boldsymbol{k}_{\parallel})$,
with $k_{\perp}$ being perpendicular to the BIS and pointing to the region $\bar{\mathcal{V}}_{\mathrm{BIS}}$ where $h_{0}(\mathbf{k})>0$.

Note that the SO field opens the gap at the BISs. We shall show that the SO field on the BISs determines the topology
uniquely. To prove it, we take advantage of the fact that the topology remains
unchanged under any continuous deformation that does not close the gap~\cite{Chan2017-2}. Therefore,
we can keep the SO field only in the region near the BISs but turn it off
elsewhere, which results in the following deformed Hamiltonian
\begin{equation}\label{eq:s_deformedH}
\mathcal{H}(\mathbf{k})\rightarrow\vec{h}'(\mathbf{k})\cdot\vec{\gamma}=\sin\Theta_{\mathbf{k}}\gamma_{0}+\cos\Theta_{\mathbf{k}}\sum_{i=1}^{d}\hat{h}_{\mathrm{so},i}(\mathbf{k})\gamma_{i},
\end{equation}
where $\hat{\boldsymbol{h}}_{\mathrm{so}}(\mathbf{k})\equiv\boldsymbol{h}_{\mathrm{so}}(\mathbf{k})/|\boldsymbol{h}_{\mathrm{so}}(\mathbf{k})|$ denotes the unit SO field. Here $\Theta_{\mathbf{k}}\in[-\pi/2,\pi/2]$ is a smooth and
monotonic function of $h_{0}(\mathbf{k})$, which approaches the step function and maps
$h_{0}(\mathbf{k})$ to $\pm\pi/2$ if $h_{0}(\mathbf{k})\gtrless0$
or to $0$ if $h_{0}(\mathbf{k})=0$, see Fig.~\ref{fig:fig5}d.
For example, it can be chosen as the error function $\pi\mathrm{erf}(\xi h_{0}(\mathbf{k}))/2$
with $\xi\to\infty$. On the other hand, the
SO field with pre-factor $\cos\Theta_{\mathbf{k}}$ is truncated to the region near the BISs. It is noteworthy
that this deformation keeps the information of BISs including the SO field on
it, and the deformed Hamiltonian has been normalized.

The bulk-surface duality can be explicitly proven with the deformed Hamiltonian~(\ref{eq:s_deformedH}).
The proof is given as follows. Generally, the topological phases in $d=2n-1$ dimensions are characterized
by the $\left(2n-1\right)$D winding number \cite{Shiozaki2014}
\begin{equation}
{\cal N}_{2n-1}=\frac{\left(-1\right)^{n-1}\left(n-1\right)!}{2\left(2\pi\ui\right)^{n}\left(2n-1\right)!}\int_{{\rm BZ}}{\rm Tr}\big[\gamma\mathcal{H}\left(\ud\mathcal{H}\right)^{2n-1}\big],
\end{equation}
where $\gamma=(\ui)^{n}\prod^{2n-1}_{i=0}\gamma_{i}$ is the chiral symmetry for the Hamiltonian, satisfying $\gamma^{\dagger}=\gamma$, $\gamma^{2}=\mathbbm{1}$ and $\{\gamma,\gamma_{i}\}=0$.
By using the deformed Hamiltonian~(\ref{eq:s_deformedH}), we obtain
\begin{widetext}
\begin{equation}
{\rm Tr}\big[\gamma\mathcal{H}\left(\ud\mathcal{H}\right)^{2n-1}\big]=\left(2n-1\right)!\left(-2\ui\right)^{n}\sum_{i=0}^{2n-1}\epsilon^{i01\cdots\bar{i}\cdots\left(2n-1\right)}h'_{i}\ud h'_{0}\wedge\ud h'_{1}\wedge\cdots\wedge\ud h'_{\bar{i}}\wedge\cdots\wedge\ud h'_{2n-1},
\end{equation}
\end{widetext}
where `d' denotes the exterior derivative, $\bar{i}$ means that the index does not take the value $i$. Here the trace property of Clifford matrices ${\rm Tr}\left[\gamma\gamma_{i}\gamma_{j}\cdots\gamma_{s}\gamma_{t}\right]=\left(-2\ui\right)^{n}\epsilon^{ij\cdots st}$ for dimensions $d=2n-1$ (see Supplementary Material), contributes the factor $(-2\ui)^{n}$, while the product of the Levi-Civita tensor and the wedge will lead to the term $(2n-1)!\epsilon^{i01\cdots\bar{i}\cdots\left(2n-1\right)}=(2n-1)!(-1)^{i}$. Similarly, in $d=2n$ dimensions, via the the trace property of Clifford matrices ${\rm Tr}\left[\gamma_{i}\gamma_{j}\cdots\gamma_{s}\gamma_{t}\right]=\left(-2\ui\right)^{n}\epsilon^{ij\cdots st}$ (see Supplementary Material), the topological phases are characterized by the $d$D (or the $n$-th) Chern number \cite{Shiozaki2014}
\begin{equation}
{\rm Ch}_{n}=-\frac{1}{2^{2n+1}}\frac{1}{n!}\left(\frac{\ui}{2\pi}\right)^{n}\int_{{\rm BZ}}{\rm Tr}\big[\mathcal{H}\left(\ud\mathcal{H}\right)^{2n}\big],
\end{equation}
with
\begin{widetext}
\begin{equation}
{\rm Tr}\big[\mathcal{H}\left(\ud\mathcal{H}\right)^{2n}\big]=\left(2n\right)!\left(-2\ui\right)^{n}\sum_{i=0}^{2n}\epsilon^{i01\cdots\bar{i}\cdots\left(2n\right)}h'_{i}\ud h'_{0}\wedge\ud h'_{1}\wedge\cdots\wedge\ud h'_{\bar{i}}\wedge\cdots\wedge\ud h'_{2n}.
\end{equation}
From the above result, one can observe that the winding number and Chern number have the same structure and are proportional to the terms $\sum^{d}_{i=0}(-1)^{i}h'_{i}\bar{\bigwedge}^{d}_{l=0}\ud h'_{l}$, where the {\em wedge product} $\bar{\bigwedge}$ indicates the index $l\neq i$. It is therefore convenient to consider these terms at first, which can be readily calculated with the deformed Hamiltonian (\ref{eq:s_deformedH}):
\begin{equation}\label{dDimension2}
\sum^{d}_{i=0}(-1)^{i}h'_{i}\bar{\bigwedge}^{d}_{l=0}\ud h'_{l}=\sin\Theta_{\mathbf{k}}\cos^{d}\Theta_{\mathbf{k}}\bigwedge_{l=1}^{d}\ud\hat{h}_{\mathrm{so},l}+\cos^{d-1}\Theta_{\mathbf{k}}\ud\Theta_{\mathbf{k}}\wedge\sum_{i=1}^{d}(-1)^{i}\hat{h}_{\mathrm{so},i}\bar{\bigwedge}_{l=1}^{d}\ud\hat{h}_{\mathrm{so},l}.
\end{equation}
\end{widetext}
To further simplify the above formula, one can choose a special configuration for the $\Theta_{\bold k}$ function, while the final result is independent of the choice of $\Theta_{\bold k}$. For this we let the SO field ($\propto\cos\Theta_{\mathbf{k}}$) for the deformed Hamiltonian be truncated to the BISs ($\sin\Theta_{\mathbf{k}}=0$) by taking $\Theta_{\mathbf{k}}$ to
approach the step function. Then the contribution to the integral from the first term on the r.h.s. of Eq.~(\ref{dDimension2}) vanishes. On the other hand, since $\ud\Theta_{\mathbf{k}}$ tends to be a delta function, the whole integral can be reduced to that over the region $\mathcal{R}$ with non-zero $\ud\Theta_{\mathbf{k}}$ (see Fig.~\ref{fig:fig5}c-d). By decomposing the measure $\ud^{d}\mathbf{k}$ into $\ud k_{\perp}\ud^{d-1}\boldsymbol{k}_{\parallel}$ (see Fig.~\ref{fig:fig5}c), the integral for Eq.~(\ref{dDimension2}) over the whole BZ can further reduce to the integral over the BISs, as shown by
\begin{align}
&\int_{\mathrm{BZ}}\sum^{d}_{i=0}(-1)^{i}h'_{i}\bar{\bigwedge}^{d}_{l=0}\ud h'_{l} \nonumber\\
= & \int_{\mathcal{R}}\cos^{d-1}\Theta_{\mathbf{k}}\ud\Theta_{\mathbf{k}}\wedge\sum_{i=1}^{d}(-1)^{i}\hat{h}_{\mathrm{so},i}\bar{\bigwedge}_{l=1}^{d}\ud\hat{h}_{\mathrm{so},l}\nonumber  \\
= & \int_{-\pi/2}^{\pi/2}\ud\Theta_{k_{\perp}}\cos^{d-1}\Theta_{k_{\perp}} \nonumber \\
  & \qquad\times\sum_{j}\int_{{\rm BIS}_{j}}\sum_{i=1}^{d}(-1)^{i}\hat{h}_{\mathrm{so},i}\bar{\bigwedge}_{l=1}^{d}\ud\hat{h}_{\mathrm{so},l}.
\end{align}

Finally, by a straightforward calculation of the integral $\int^{\pi/2}_{-\pi/2}\ud x\,\cos^{d-1}x$, both the winding number and Chern number can be expressed in a unified form as the winding of the SO field $\hat{\boldsymbol{h}}_{\mathrm{so}}$ along the BISs:
\begin{equation}
w_{d-1}=\sum_{j}\frac{\Gamma(d/2)}{2\pi^{d/2}}\frac{1}{\left(d-1\right)!}\int_{{\rm BIS}_{j}}\hat{\boldsymbol{h}}_{\mathrm{so}}\bigl(\ud\hat{\boldsymbol{h}}_{\mathrm{so}}\bigr)^{d-1},\label{eq:s_socwinding}
\end{equation}
where $\hat{\boldsymbol{h}}_{\mathrm{so}}\bigl(\ud\hat{\boldsymbol{h}}_{\mathrm{so}}\bigr)^{d-1}\equiv\epsilon^{i_{1}i_{2}\cdots i_{d}}\hat{h}_{\mathrm{so},i_{1}}\ud\hat{h}_{\mathrm{so},i_{2}}\wedge\cdots\wedge\ud\hat{h}_{\mathrm{so},i_{d}}$ is defined on the $(d-1)$D BISs, with the indices $i_{1,\dots,d}\in\{1,\dots,d\}$, and $\Gamma(x)$ is the Gamma function with $\Gamma(x+1)=x\Gamma(x)$,
$\Gamma(1)=1$ and $\Gamma(1/2)=\sqrt{\pi}$.
Note that the prefactor $2\pi^{d/2}/\Gamma(d/2)$ is actually the area of a $(d-1)$D unit sphere.
Analogous to the ``Gauss's law'', the invariant $w_{d-1}$ simply counts the total topological charges of the SO field located at momenta with $\boldsymbol{h}_{\mathrm{so}}(\bold k)=0$ and enclosed by the BISs. Finally we reach
\begin{equation}
\nu_d=w_{d-1}=\sum_{i\in \mathcal{V}_{\mathrm{BIS}}}\mathcal{C}_{i},\label{eq:s_charges}
\end{equation}
where
\begin{equation}
\mathcal{C}_{i}\equiv\frac{\Gamma(d/2)}{2\pi^{d/2}}\frac{1}{\left(d-1\right)!}\int_{\mathrm{BIS}_{i}}\hat{\boldsymbol{h}}_{\rm so}\bigl(\ud\hat{\boldsymbol{h}}_{\rm so}\bigr)^{d-1}
\end{equation}
represents the total topological charge in the patch of ${\cal V}_{\rm BIS}$ enclosed by the $i$-th BIS. The topological phase transition occurs when a charge passes through a BIS, since the gap must be closed and reopen as the topological charge located at $\boldsymbol{h}_{\mathrm{so}}({\bf k})=0$ passes through the BISs with  $h_{0}({\bf k})=0$. Accordingly, the topology characterized by the SO field on BISs is changed.

\begin{figure*}[htpb]
\includegraphics{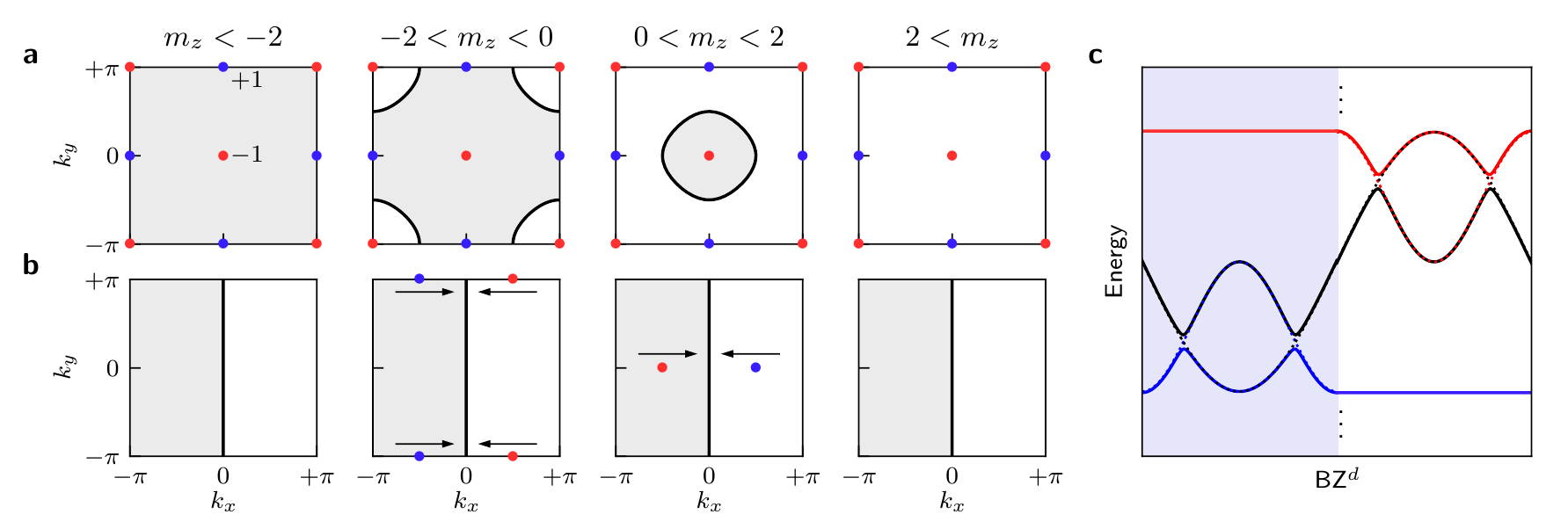}
\caption{\textsf{\textbf{Charge configurations of a 2D model and schematic diagram of the multiband case}}. \textsf{\textbf{a-b}}, The topological charges by taking that $h_{0}=m_{z}-t_{0}\cos k_{x}-t_{0}\cos k_{y}$ (\textsf{\textbf{a}})
and $h_{0}=t_{\mathrm{so}}\sin k_{x}$ (\textsf{\textbf{b}}), respectively. The red (blue) circle
denotes the charge $-1$ ($+1$),
and the gray region represents $h_{0}(\mathbf{k})<0$ (i.e., $\mathcal{V}_{\mathrm{BIS}}$) with black lines being BISs.
\textsf{\textbf{c}}, A schematic diagram for multiple bands with (solid lines) or without (dotted lines) the SO field.
The different background colors indicate different patches.}
\label{fig:fig6}
\end{figure*}

To gain intuition, we take the Chern insulator as an example, with Hamiltonian given by $\mathcal{H}(\mathbf{k})=\vec{h}(\mathbf{k})\cdot\vec{\sigma}$ with $\vec{h}(\mathbf{k})=(t_{\mathrm{so}}\sin k_{x},t_{\mathrm{so}}\sin k_{y}, m_{z}-t_{0}\cos k_{x}-t_{0}\cos k_{y})$. According to the bulk-surface duality, the Chern number reduces to the form $\mathrm{Ch}_{1}=\sum_{j}\frac{1}{2\pi}\epsilon^{mn}\oint_{\mathrm{BIS}_{j}}\ud\hat{\ell}\cdot\hat{{h}}_{\mathrm{so},m}\nabla\hat{h}_{\mathrm{so},n}$,
counting the topological charges of the SO field $\boldsymbol{h}_{\mathrm{so}}$ within ${\cal V}_{\rm BIS}$~\cite{SI}.
First, we take that $h_{0}\equiv h_{z}$ and $\boldsymbol{h}_\mathrm{{so}}\equiv (h_{y},h_{x})$. The topological charges and BISs are shown in Fig.~\ref{fig:fig6}a, which explicitly determine Chern number for different phases governed by $m_{z}$: The Chern number $\rm{Ch}_{1}=-1$ for $0<m_{z}<2$; $\rm{Ch}_{1}=+1$ for $-2<m_{z}<0$; $\rm{Ch}_{1}=0$ for otherwise.
Further, if we choose $h_{0}\equiv h_x$ such that $\mathcal{V}_{\mathrm{BIS}}$
with $h_{0}(\mathbf{k})<0$ corresponds to the region with $-\pi<k_{x}<0$, and then
the SO field is $\boldsymbol{h}_{\mathrm{so}}\equiv(h_z,h_y)$.
Two topological charges with opposite signs, located at $\mathbf{k}=\left(\pm\arccos[\left(m_{z}+1\right)/t_{0}],\pi\right)$ for $-2<m_z<0$
or at $\bold k=\left(\pm\arccos[(m_{z}-1)/t_{0}],0\right)$ for $0<m_z<2$, are created from vacuum,
moves and finally annihilates, as $m_{z}$ varies from $-\infty$ to
$+\infty$ [see Fig.~\ref{fig:fig6}b], which characterizes the same Chern number for each topological regime as in Fig.~\ref{fig:fig6}a.

\subsection{2. Generalization to generic multiband systems}

We show here that the bulk-surface duality is valid for the generic systems with arbitrary multiple
bands and multiple band crossings in the BZ, as illustrated Fig.~\ref{fig:fig6}c. {Note that for a generic $d$D topological phase, the gap must be opened through each group of $n_d$ bands ($n_d=2^{d/2}$ or $2^{(d+1)/2}$) with hybridization at BISs of these bands.
Therefore, it is more convenient, by transforming the system into a new trivial bases, which block diagonalize the Hamiltonian $\mathcal{H}(\mathbf{k})=\mathcal{H}_{1}(\mathbf{k})\oplus\mathcal{H}_{2}(\mathbf{k})\oplus\cdots$ at each momentum $\mathbf{k}$. Only those blocks involving $n_{d}$ bands with band crossings have nontrivial contribution to the topology. Note that this transformation does not change the topology of the Block Hamiltonians due to the triviality of the new bases.}
For convenience, we denote each band crossing for a group of $n_d$ bands as a BIS. Without changing the topology of the system, we assume that each BIS is separated from others in the BZ (If there are more than one BIS occurring at the same $\bold k$ point, one can continuously deform the band dispersion and separate such BISs without affecting the bulk topology). In this case, the Hamiltonian around each BIS can be decomposed to the contributions from three parts: the $n_d$ bands with hybridization on the BIS, the bands with energy above the BIS, and the bands below the BIS. Thus the Hamiltonian can be locally written as $\mathcal{H}(\mathbf{k})=\mathcal{P}_{\mathrm{upper}}(\mathbf{k})\oplus\mathcal{H}_{\text{BIS}}(\mathbf{k}) \oplus\mathcal{P}_{\mathrm{lower}}(\mathbf{k})$.
Here $\mathcal{P}_{\mathrm{upper/lower}}$ are the projection
operators for the bands with higher (lower) energy than the part with
band crossing $\mathcal{H}_{\text{BIS}}$. For the higher or lower bands $\mathcal{P}_{\mathrm{upper/lower}}$, as there is no band crossing, the corresponding spin textures are trivial and no singularities exist for
the corresponding eigenstates. One can safely deform the bands of $\mathcal{P}_{\mathrm{upper/lower}}$
into constants in dispersion. Similarly, away from the BISs, no band crossing occurs, so all the bands can be smoothly connected in those areas of the BZ, and also be further deformed into constants in dispersion. It is straightforward to know that the flattened pathes in BZ have no contribution to the bulk topology, which is then solely determined from the integral over the BISs. Further, to compute the contribution from a specific BIS, we can single out the $n_d$ bands which are characterized by the Hamiltonian $\mathcal{H}_{\text{BIS}}(\mathbf{k})$ on the BIS, and flatten these bands in all the remaining $\bold k$ space of BZ away from such BIS. This can be safely done since the band crossings occurring at other momenta of BZ are unrelated to the contribution to bulk topology from the BIS under consideration (thus those band crossings can be removed for this computation). In this way, the contribution of the BIS is effectively determined by a basic Hamiltonian with the band crossing at BIS described by $\mathcal{H}_{\text{BIS}}(\mathbf{k})$, precisely given by the $(d-1)$D topological number defined on the BIS. The total topological invariant of the $d$D phase is then a summation of contributions from all BISs, formed by all bands of the BZ.
This completes the proof.

The above proof implies that the $(d-1)$D topological number shown in Eq.~\eqref{eq:s_socwinding} is also valid for the case with multiple bands. However, in this generic case the SO field $\boldsymbol{h}_{\rm so} (\bold k)$ is defined locally on each BIS, characterizing only the hybridization among the $n_d$ bands of such BIS.

\section{Appendix B: Dynamical Bulk-Surface Correspondence}\label{sec2}
\renewcommand{\theequation}{B\arabic{equation}}
\setcounter{equation}{0}  

In this section, we present the details of the dynamical bulk-surface correspondence.
Two basic issues need to be addressed: how to identify the BISs and how to extract the dynamical topological invariant.
The study is first based on the basic Hamiltonian (\ref{model}), and then generalized to generic multiband systems.

\subsection{1. Characterizing the BISs by quench dynamics}

As discussed in the main text, the
system is initially prepared in the ground state of the pre-quench Hamiltonian
$\mathcal{H}(m_{0}\vert_{t<0},\mathbf{k})$ in the deep trivial regime, with density matrix $\rho_{0}$, where the spins are
fully polarized in the opposite direction of the axes $h_{0}$.
After a sudden quench to a topological
non-trivial Hamiltonian $\mathcal{H}(m_{0}\vert_{t>0},\mathbf{k})$, each momentum-linked spin
will rotate around the post-quench vector field $\vec{h}(\mathbf{k})$, with the quantum dynamics being governed by the unitary evolution operator $U(t)=\exp(-\ui\mathcal{H}t)$ for the post-quench system, see Fig.~\ref{fig1}c.


We consider the time-averaged spin polarizations, which can be measured in experiments:
\begin{equation}
\overline{\left\langle \gamma_{i}\right\rangle }=\lim_{T\to\infty}\frac{1}{T}\int_{0}^{T}\ud t\,\mathrm{Tr}\left[\rho_{0} e^{\ui\mathcal{H}t}\gamma_{i}e^{-\ui\mathcal{H}t}\right].
\end{equation}
For the Hamiltonian (\ref{model}), one can show that $e^{\ui\mathcal{H}t}=\cos(Et)+\ui\sin(Et)\mathcal{H}/E$ with $E(\bold k)=\sqrt{\sum_{i=0}^{d}h_{i}^{2}}$ the energy. Thus the time-averaged spin polarizations read
\begin{equation}
\overline{\left\langle \gamma_{i}(\bold k)\right\rangle }=h_{i}\mathrm{Tr}\left[\rho_{0}\mathcal{H}\right]/E^{2}=-h_{i}(\bold k)h_{0}(\bold k)/E^{2}(\bold k).
\end{equation}
On the BISs with $h_{0}(\mathbf{k})=0$ of the post-quench Hamitlonian, one can find that all the time-averaged spin polarizations $\overline{\left\langle \gamma_{i}(\bold k)\right\rangle }$ (for $i=0,1,...,d$) vanish on the BISs.
Hence, the BISs can be dynamically determined by the surface with vanishing time-averaged spin polarizations:
\begin{equation}
\mathrm{BIS}=\{\mathbf{k}\vert\overline{\left\langle\vec{\gamma}(\mathbf{k})\right\rangle}=0\}.
\end{equation}

\subsection{2. Dynamical topological invariants and the dynamical bulk-surface correspondence}

{Time-averaged spin polarizations $\overline{\left\langle \gamma_{i}\right\rangle }$ vanish on the BISs, but are nonzero inside and outside the BISs. The variation of $\overline{\left\langle \gamma_{i}\right\rangle }$ across the BISs can characterize the topology. For this a new dynamical spin-texture field is defined in the main text as the directional derivative of time-averaged spin textures} $\overline{\left\langle\vec{\gamma}\right\rangle }$ across the BISs:
\begin{align}
\widetilde{{\bold g}(\mathbf{k})} & \equiv-\frac{1}{{\cal N}_{\bold k}}\partial_{k_{\perp}}\overline{\left\langle\vec{\gamma}\right\rangle }=-\lim_{k_{\perp}\to 0^{+}}\frac{1}{{\cal N}_{\bold k}}\frac{\Delta\overline{\left\langle\vec{\gamma}\right\rangle }}{2k_{\perp}} \nonumber\\
&=-\lim_{k_{\perp}\to 0^{+}}\frac{1}{{\cal N}_{\bold k}}\frac{\overline{\left\langle \vec{\gamma}\right\rangle }\vert_{h_{0}(k_{\perp})>0}-\overline{\left\langle\vec{\gamma}\right\rangle }\vert_{h_{0}(-k_{\perp})<0}}{2k_{\perp}},
\end{align}
where ${\cal N}_{\bold k}$ is the normalization factor, and the origin of $k_{\perp}$ has been shifted locally to the BISs.
Note that for the pre-quench Hamiltonian with $h_{0}(\mathbf{k})\ll 0$, the time averaged spin polarizations $\overline{\left\langle \gamma_{i}\right\rangle }=+h_{i}h_{0}/E^{2}$. Accordingly, the dynamical spin-texture field should be defined by $\widetilde{{\bold g}(\mathbf{k})}\simeq\partial_{k_{\perp}}\overline{\left\langle\vec{\gamma}\right\rangle }$, which takes an additional minus sign.

Near the BISs, the dispersion $h_{0}(\mathbf{k})$ is monotonous and is determined by $k_{\perp}$ uniquely. One can identify $h_{0}$ from $k_{\perp}$, that is, $h_{0}\cong k_{\perp}$ up to some local scale transformation that does not affect the topology. The difference of $\overline{\left\langle \gamma_{i}\right\rangle }$ between the two sides of BISs reads
\begin{eqnarray}
\Delta\overline{\left\langle \gamma_{i}\right\rangle }\bigr|_{k_{\perp}\rightarrow0}&=& - \biggr[\frac{h_{i}(k_{\perp},\boldsymbol{k}_{\parallel})k_{\perp}}{k_{\perp}^{2}+\sum_{j=1}^{d}h_{j}^{2}(k_{\perp},\boldsymbol{k}_{\parallel})}-\nonumber\\
&&-\frac{h_{i}(-k_{\perp},\boldsymbol{k}_{\parallel})\left(-k_{\perp}\right)}{k_{\perp}^{2}+\sum_{j=1}^{d}h_{j}^{2}(-k_{\perp},\boldsymbol{k}_{\parallel})}\biggr]\nonumber\\
&=&- 2\frac{h_{i}(0,\boldsymbol{k}_{\parallel})k_{\perp}}{\sum_{j=1}^{d}h_{j}^{2}(0,\boldsymbol{k}_{\parallel})}+\mathcal{O}(k_{\perp}^{2}).
\end{eqnarray}
We obtain the emergent dynamical spin-texture field by
\begin{equation}
\widetilde{g_{i}(\mathbf{k})}\bigr|_{\bold k\in\rm BISs}=\frac{h_{i}(0,\boldsymbol{k}_{\parallel})}{\sum_{j=1}^{d}h_{j}^{2}(0,\boldsymbol{k}_{\parallel})}=\hat{h}_{\mathrm{so},i}.
\end{equation}
With this result we reach that the bulk topology can be characterized by the dynamical invariant defined as the winding of the new dynamical field $\widetilde{{\bold g}(\mathbf{k})}$ on the BISs:
\begin{equation}\label{dynamicalSM1}
w_{d-1}\equiv\sum_{j}\frac{\Gamma(d/2)}{2\pi^{d/2}}\frac{1}{\left(d-1\right)!}\int_{{\rm BIS}_{j}}\widetilde{{\bold g}(\mathbf{k})}\big[\ud\widetilde{{\bold g}(\mathbf{k})}\big]^{d-1}.
\end{equation}
This is the dynamical bulk-surface correspondence.

\subsection{3. Quench dynamics in multi-band systems}

{Now we show the dynamical classification for multi-band systems. To better understand our results, we consider first the weak spin-orbit coupling regime, where the band dispersions are dominated by the bare energies $\epsilon_j({\bf k})$ of each bare orbital $\lambda_j$ which define the BISs, and the spin-orbit couplings are weak and induce small gap on the BISs. Then we consider the generic case, including the strong spin-orbit coupling regime, namely, the spin-orbit couplings can even dominate the dispersions of the topological bands. For convenience, we consider here the 1D or 2D topological phases, which are most relevant for the current cold atom experiments, and the proof can be directly applied to the higher dimensional phases.

For the weak coupling regime, the spin-orbit couplings are weak, so that they only have local effects in the
momentum space and cannot change the band structure determined by the bare energies of orbitals dramatically. Note that for a 1D/2D system, a key feature is that the topological gap must be opened between every two bands. Therefore, all one needs to determine is the information for all these band crossings determined by these bare orbitals, namely, the BISs defined between every two bands. Then the detection scheme for the minimal Hamiltonian can be generalized directly. In
particular, we initialize the system in the deep trivial regime, with all subbands for orbitals being far separated from each other (similar to the large Zeeman splitting in the minimal model) and prepare the system in a single band (denoted as $\lambda$). Then we pick up this band and another one $\lambda_1$, and quench the effective Zeeman splitting $m_1$ between the two bands (the energy splitting between them) to desired post-quench regime, but the parameters for all the remaining bands keep unchanged. With this process we can dynamically determine the BISs defined by $\lambda$ and $\lambda_1$ bands, and the topological invariants defined on the BISs, by measuring the Pauli operators defined in these two orbitals. Then, replacing $\lambda_1$ with another band $\lambda_2$ and through a similar process we can detect the topology on the BISs defined by $\lambda$ and $\lambda_2$ bands. With $\lambda_j$ running over all the bands, we obtain the topology of $\lambda$ band by summing over the invariants detected in these processes. Repeating this detection for all $\lambda$ bands we obtain the topology for all subbands of the whole system. It is trivial to know that when the band number is two, a single quench determines the topology, reducing to the minimal model regime.

Now we show the detection scheme for the generic case without assuming the spin-orbit coupling to be weak. The key difference between the weak and strong spin-orbit coupling regimes is that, in the strong coupling regime the band crossing between two subbands can be affected by the couplings between such two bands and the remaining bands. Thus picking up only every two subbands for quench may not extract the true topology. Nevertheless our detection scheme can be still applied. Again, the key observation is that the topological gap has to be opened between every two bands (for 1D/2D system). Therefore, at every momentum $\bold k$, by transforming the system into a new trivial bases, we can always block diagonalize the Hamiltonian $\mathcal{H}(\bold k)=\mathcal{H}_1(\mathbf{k})\oplus \mathcal{H}_2(\mathbf{k})\oplus\cdots$, with each block $\mathcal{H}_j$ being at most two by two matrix.  More generically, this block diagonal picture is necessary only in the vicinity of each BIS.
Note that all the band crossing information can be kept in each block diagonalized Hamiltonian $\mathcal{H}_j$, and only these blocks with band crossings have nontrivial contribution to the topology, with the topology determined by the corresponding BISs defined by $h_0^{(j)}$ axis in $\mathcal{H}_j$. All we need to do is detect the topological invariants on these BISs. However, in the present block diagonal picture, the (pseudo)spin-$1/2$ bases for each block Hamiltonian $\mathcal{H}_j$ is in general momentum dependent, rather than being constant, while such bases do not have nontrivial topology. This is the key difference from the minimal model and weak coupling regime. The dynamical detection of the topology can be performed in the following way. We initialize the system in the deep trivial regime with the energies of the bare orbitals being far separated from each other. The bare orbitals are again denoted as $\lambda_j$, and note that they are momentum-independent. The (pseudo)spin bases of each block Hamiltonian $\mathcal{H}_j$, denoted by $\uparrow^{(j)}_{\bold k}$ and $\downarrow^{(j)}_{\bold k}$, are obtained by momentum-dependent unitary transformations on the bare orbitals. The system is prepared in certain orbital $\lambda$, and is then quenched to the final total Hamiltonian $\mathcal{H}$. For each block $\mathcal{H}_j$, the initial state is simply obtained by projecting $\lambda$-state into the (pseudo)spin bases of $\mathcal{H}_j$, i.e., $|0_{\bold k}\rangle=(c_{\uparrow}(\mathbf{k}),c_{\downarrow}(\mathbf{k}))^{\mathsf{T}}$, which is momentum-dependent. This is equivalent to the case that for the minimal model, the initial trivial state is not fully polarized, but momentum-dependent. Accordingly, in quench dynamics, we can measure the momentum-dependent observable $\sigma^{(j)}_{z;\bold k}=|0_{\bold k}\rangle\langle 0_{\bold k}|-|0'_{\bold k}\rangle\langle 0'_{\bold k}|$, where $|0'_{\bold k}\rangle$ is the state perpendicular to $|0_{\bold k}\rangle$, and of which the time-averaged expectation value reads
\begin{align}
\overline{\langle \sigma^{(j)}_{z;\bold k}\rangle} & =\lim_{T\to\infty}\frac{1}{T}\int^{T}_{0}\ud t\,\mathrm{Tr}(|0_{\bold k}\rangle\langle 0_{\bold k}|e^{\ui\mathcal{H}_{j}(\bold k)t}\sigma^{(j)}_{z;\bold k}e^{-\ui\mathcal{H}_{j}(\bold k)t}) \nonumber\\
& = \lim_{T\to\infty}\frac{1}{T}\int^{T}_{0}\ud t\,\mathrm{Tr}(\left|0\right\rangle\left\langle 0\right|e^{\ui\mathcal{H}'_{j}(\bold k)t}\sigma^{(j)}_{z}e^{-\ui\mathcal{H}'_{j}(\bold k)t}),
\end{align}
where the second line is obtained by a unitary transformation $U(\mathbf{k})=(|0_{\bold k}\rangle,|0'_{\bold k}\rangle)$, and the Hamiltonian $\mathcal{H}'_{j}(\mathbf{k})=U(\bold k)\mathcal{H}_{j}(\mathbf{k})U^{\dagger}(\bold k)$ is topologically equivalent to $\mathcal{H}_{j}$ since $|0_{\bold k}\rangle$ and $|0'_{\bold k}\rangle$ are trivial. Here $\left|0\right\rangle=(1,0)^{\mathsf{T}}$ and $\sigma^{(j)}_{z}$ is the usual Pauli matrix. The above equality means that one has chosen a momentum-dependent $\sigma^{(j)}_{z;\bold k}$ axis to define the BISs for $\mathcal{H}_j$ block, which coincide with the BISs for the Hamiltonian $\mathcal{H}'_{j}$ with the same topology as $\mathcal{H}_{j}$. Therefore, this momentum-dependent $h^{(j)}_{0}$ axis does not affect the existence of BISs, nor the topology defined on such BISs.
Then, as long as the projection of $\lambda$-state to the (pseudo)spin bases of $\mathcal{H}_j$ is nonzero, from the (pseudo)spin dynamics one can extract the dynamical topological invariant for the block Hamiltonian $\mathcal{H}_j$. Further, by initializing the system in different bare orbital $\lambda$-states and performing quench until the projection of initial states can cover (pseudo)spin bases of all Hamiltonian blocks, one can obtain the topological invariants for all $\mathcal{H}_j$ blocks, which characterize the topology of all bands of the total Hamiltonian.

\begin{figure}
\includegraphics[width=\columnwidth]{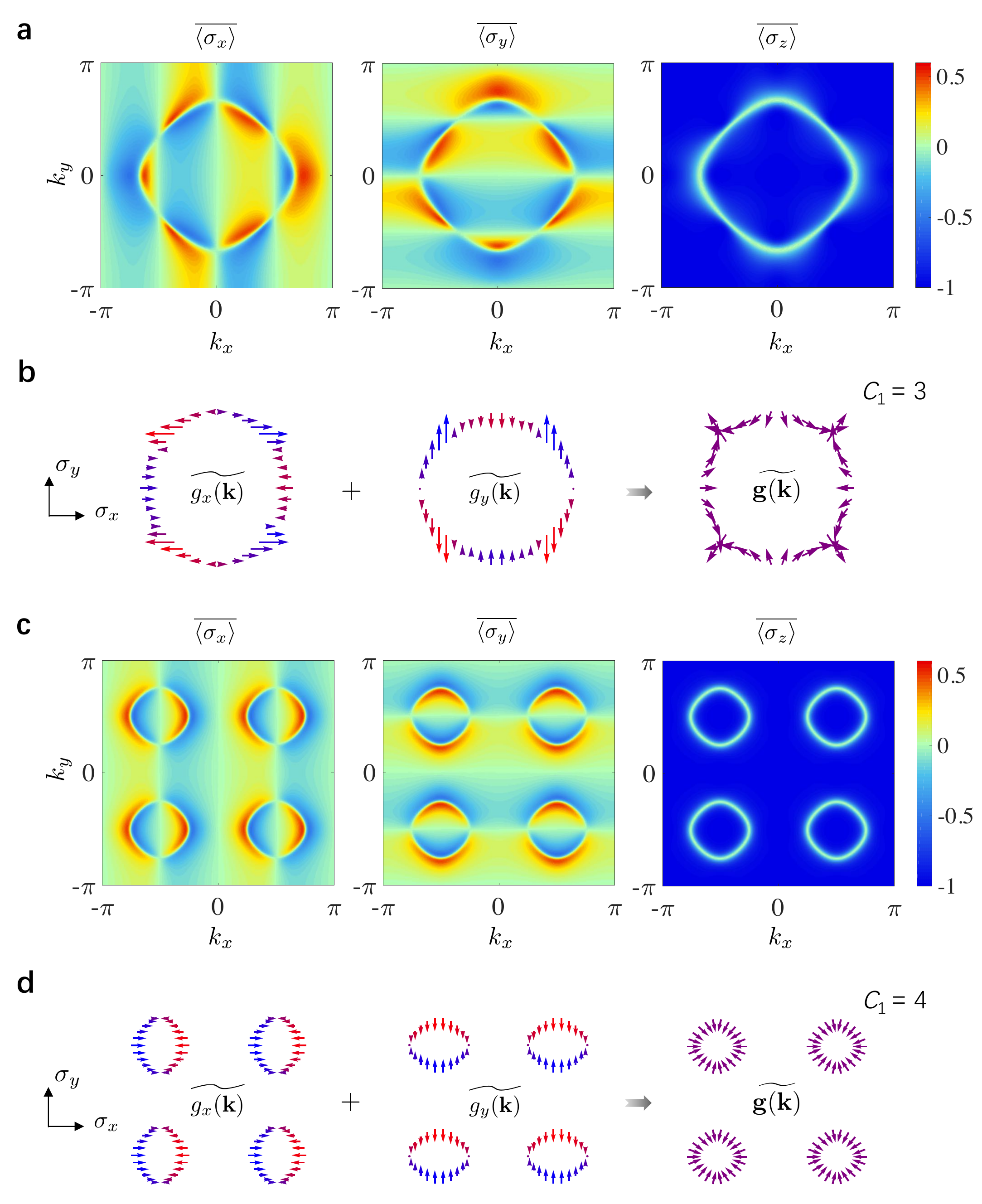}
\caption{\textsf{\textbf{2D topological phases with high Chern numbers}}.
\textsf{\textbf{a-b}}, Dynamical results of the model ${\cal H}_1$ in section III(B).
Time-averaged spin textures ({\bf{\textsf a}}) also exhibit a ring structure for the BIS. But the emergent dynamical field ({\bf{\textsf b}}) winds three circles, corresponding to $C_1=3$. The results are obtained by a quench from $m_z=8t_0$ to $0.5t_0$ with $t_{\rm so}=0.2t_0$.
\textsf{\textbf{c-d}}, Dynamical results of the model ${\cal H}_2$ of section III(B).
Time-averaged spin textures ({\bf{\textsf c}}) show four rings, and each contributes a full winding ({\bf{\textsf d}}), charactering the Chern number $C_1=4$.
The quench is taken from $m_z=8t_0$ to $-t_0$ with $t_{\rm so}=0.2t_0$.
}\label{fig:fig7}
\end{figure}

\begin{figure*}
\includegraphics[width=0.75\textwidth]{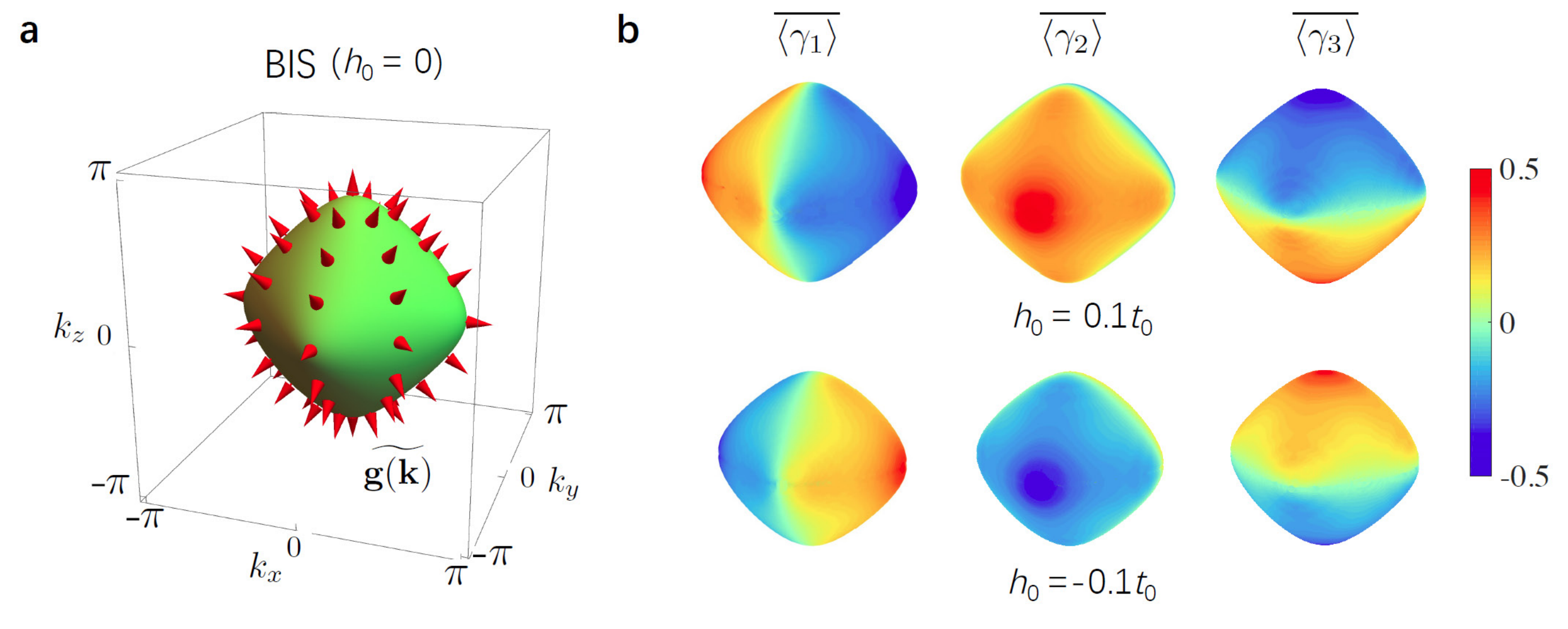}
\caption{\textsf{\textbf{Numerical results of the 3D model}}.
The 2D BIS of the topological phase with 3D winding number $\nu_3=-1$ given from the dynamical field $\widetilde{\boldsymbol{g}(\mathbf{k})}$ ({\bf{\textsf a}}). The dynamical field $\widetilde{\boldsymbol{g}(\mathbf{k})}=-\partial_{k_{\perp}}\overline{\langle\vec{\gamma}\rangle}$ is qualitatively the same as that obtained from the difference of the time-averaged spin textures $\overline{\langle\gamma_j\rangle}$ ($j=1,2,3$) on two equal-energy surfaces $h_0=\pm0.1 t_0$ ({\bf{\textsf b}}). The results are obtained by spin dynamics of two orbitals after a quench from $m_z=8t_0$ to $1.3t_0$, with $t_{\rm so}=0.2t_0$.
}\label{fig:fig8}
\end{figure*}

We note that in the real cold atom experiment, one can always simplify the measurement by well controlling the parameters, e.g. staring from the weak spin-orbit regime. In this way, one can get enough information of the bands and topology before entering the strong coupling regime, and simplify the detection in generic case. Therefore, our dynamical classification theory is universal for the case of multiband systems.}

\section{Appendix C: Numerical Results}\label{sec3}

In this section, we provide more numerical calculations related to the main text for the quench study.

\subsection{1. Topological phases with high invariant}

The dynamical classification theory is clearly applicable to topological phases with high invariants. To illustrate the validity, we consider two different 2D models ${\cal H}_j({\bf k})={\vec h}_j({\bf k})\cdot{\vec \sigma}$ ($j=1,2$), which are variations of the 2D QAH model studied in the main text.

We consider the first model ${\cal H}_1$ with the vector field ${\vec h}_1({\bf k})=(t_{\rm so
}\sin 2k_x,t_{\rm so}\sin2k_y,m_z-t_0\cos k_x-t_0\cos k_y)$. While the trivial phase corresponds to $|m_{z}|>2t_{0}$, the topological phases are distinguished as: (i) $t_0<m_z<2t_0$
with the Chern number $C_1=-1$; (ii) $0<m_z<t_0$ with $C_1=3$; (iii)  $-t_0<m_z<0$ with $C_1=-3$; (iv) $-2t_0<m_z<-t_0$ with $C_1=1$.
When $t_0\gg t_{\rm so}$, this model has almost the same band structure as the 2D model in the main text.
It is then supposed that
the results of quenching to the phases with $C_1=\pm1$ should be similar to those shown in Fig.~2a. Here we only provide the numerical
results of quenching to the phase with $C_1=3$ (see Fig.~\ref{fig:fig7}a-b), where the time-averaged spin textures also exhibit ring-shape structure, but
the corresponding spin-texture field winds around three times, charactering the high Chern number.

We then consider the other one ${\cal H}_2$ with ${\vec h}_2({\bf k})=(t_{\rm so}\sin 2k_x,t_{\rm so}\sin2k_y,m_z-t_0\cos 2k_x-t_0\cos 2k_y)$. Apart from the trivial phase with $|m_{z}|>2t_{0}$, this model has two topological phases:
(i) $0<m_z<2t_0$ with the Chern number $C_1=-4$; (ii) $-2t_0<m_z<0$ with $C_1=4$. After a sudden quench from $m_z\gg 2t_0$ to $|m_z|<2t_0$, the time-averaged spin textures have four rings (see Fig.~\ref{fig:fig7}c-d), and the dynamical spin-texture field on each ring winds around once. Thus four rings in the BZ characterize the Chern number $C_1=\pm4$. {More results, including the case with a trivial phase but having multiple BISs, are discussed in Supplementary Material~\cite{SI}.}

\subsection{2. Characterizing 3D topological phases}

As discussed in the main text, the 3D model ${\cal H}={\vec h}({\bf k})\cdot{\vec \gamma}$, which is a direct generalization of the 2D model (see the main text), has three topological phases. In Fig.~\ref{fig3}, we have shown the BIS of the topological phase with 3D winding number $\nu_3=-1$ and the time-averaged spin texture
$\overline{\langle\gamma_3\rangle}$. Here we provide more results and demonstrate that the dynamical spin-texture field $\widetilde{\bold{g}(\mathbf{k})}$ on the BIS is qualitatively the same as that determined from the difference of $\overline{\langle\gamma_j\rangle}$ $(j=1,2,3)$
on the two equal-energy surfaces slightly outside ($h_0=0.1t_0$) and inside ($h_0=-0.1t_0$) the BIS (see Fig.~\ref{fig:fig8}).


\onecolumngrid

\newpage{}

\section*{\large Supplementary Material:\\Non-equilibrium classification of topological quantum phases}

\setcounter{equation}{0} \setcounter{figure}{0} \setcounter{table}{0}
\setcounter{page}{1} \makeatletter \global\long\def\theequation{S\arabic{equation}}
 \global\long\def\thefigure{S\arabic{figure}}
 
 In this Supplementary Material, we shall show the elementary representation matrices of Clifford algebra, and provide the detailed proof of bulk-surface duality in the simple cases, as well as more details of numerical results.

\section{The elementary representation matrices of Clifford algebra}

The elementary representation matrices of Clifford algebra $\{\gamma_i,\gamma_j\}=2\delta_{ij}\mathbbm{1}$ are of dimension $n_{d}=2^{d/2}$ (or $2^{(d+1)/2}$) if the dimensionality of the system $d$ is even (or odd). Here we give an example to construct these Clifford matrices and discuss their trace property, as shown in Tab.~\ref{tab:tabs1}.

\begin{table}[h]
\renewcommand{\arraystretch}{1.6}
\begin{tabular}{|c|c|c|c|}
\hline
dimension $d$ & Clifford matrices & chiral matrix & $\mathrm{Tr}[\gamma\prod^{d}_{i=0}\gamma_{i}]$ or $\mathrm{Tr}[\prod^{d}_{i=0}\gamma_{i}]$\tabularnewline
\hline
$1$ & $\gamma_{0}^{(1)}=\sigma_{x}$, $\gamma_{1}^{(1)}=\sigma_{y}$ & $\gamma^{(1)}=\ui\sigma_{x}\sigma_{y}$ & $-2\ui$\tabularnewline
\hline
$2$ & $\gamma_{0}^{(2)}=\gamma_{0}^{(1)}$, $\gamma_{1}^{(2)}=\gamma_{1}^{(1)}$,
$\gamma_{2}^{(2)}=\gamma^{(1)}$ & $-$ & $-2\ui$\tabularnewline
\hline
\multirow{2}{*}{$2n-1$} & $\gamma_{i}^{(2n-1)}=\gamma_{i}^{(2n-3)}\otimes\sigma_{z}$ for $i=0,1,\dots,2n-3$, & \multirow{2}{*}{$\gamma^{(2n-1)}=-\gamma_{2n-3}\otimes\sigma_{z}$} & $-2\ui\mathrm{Tr}[\gamma^{(2n-3)}\prod_{i=0}^{2n-3}\gamma_{i}^{(2n-3)}]$\tabularnewline
 & $\gamma_{2n-2}^{(2n-1)}=\mathbbm{1}\otimes\sigma_{x}$, $\gamma_{2n-1}^{(2n-1)}=\mathbbm{1}\otimes\sigma_{y}$ &  & $=\left(-2\ui\right)^{n}$\tabularnewline
\hline
\multirow{2}{*}{$2n$} & $\gamma_{i}^{(2n)}=\gamma_{i}^{(2n-1)}$ for $i=0,1,\dots,2n-1$, & \multirow{2}{*}{$-$} & \multirow{2}{*}{$\left(-2\ui\right)^{n}$}\tabularnewline
 &  $\gamma_{2n}^{(2n)}=\gamma^{(2n-1)}$ &  & \tabularnewline
\hline
\end{tabular}
\caption{\textbf{\textsf{Construction of Clifford matrices and their trace property}}. Here $\sigma_{x,y,z}$ are Pauli matrices. For odd dimensions $d=2n-1$, one can define a chiral matrix $\gamma=(\ui)^{n}\prod^{2n-1}_{i=0}\gamma_{i}$, which satisfies $\gamma^{\dagger}=\gamma$, $\gamma^{2}=\mathbbm{1}$ and $\{\gamma,\gamma_{i}\}=0$. These Clifford matrices have the trace property ${\rm Tr}\left[\gamma\gamma_{i}\gamma_{j}\cdots\gamma_{s}\gamma_{t}\right]=\left(-2\ui\right)^{n}\epsilon^{ij\cdots st}$ for odd dimensions, and ${\rm Tr}\left[\gamma_{i}\gamma_{j}\cdots\gamma_{s}\gamma_{t}\right]=\left(-2\ui\right)^{n}\epsilon^{ij\cdots st}$ for even dimensions, which can be obtained by induction (the last column).
\label{tab:tabs1}}
\end{table}

\section{Proof of bulk-surface duality in the simple cases}
To better understand the proof for the generic bulk-surface duality for any dimensions in the Appendix, in this section, we provide the details of the proof for the simple cases, namely, the 1D, 2D and 3D topological phases.

Below, the proof is directly based on the deformed Hamiltonian
\begin{equation}\label{eq:s_deformedH}
\mathcal{H}(\bold k)=\vec{h}'(\bold k)\cdot\vec{\gamma}=\sin\Theta_{\bold k}\gamma_{0}+\cos\Theta_{\bold k}\sum^{d}_{i=1}\hat{h}_{\mathrm{so},i}(\bold k)\gamma_{i},
\end{equation}
which is topologically equivalent to the basic Hamiltonian $\mathcal{H}(\bold k)=\vec{h}(\bold k)\cdot\vec{\gamma}$ considered in the main text. Without loss of the generality, $h_{0}$ defines the band-inversion surfaces (BISs) and $\boldsymbol{h}_{\mathrm{so}}\equiv(h_1,\dots,h_{d})$ is the spin-orbit (SO) field. Here, as discussed in the Appendix, $\Theta_{\mathbf{k}}\in[-\pi/2,\pi/2]$ is a smooth and
monotonic function of $h_{0}(\mathbf{k})$, which approaches the step function and maps
$h_{0}(\mathbf{k})$ to $\pm\pi/2$ if $h_{0}(\mathbf{k})\gtrless0$
or to $0$ if $h_{0}(\mathbf{k})=0$, see Fig.~5d in the main text.
For example, it can be chosen as the error function $\pi\mathrm{erf}(\xi h_{0}(\mathbf{k}))/2$
with $\xi\to\infty$.
Therefore, this deformed Hamiltonian extracts all important informations on the BISs, including the normalized SO field $\hat{\boldsymbol{h}}_{\mathrm{so}}=\boldsymbol{h}_{\mathrm{so}}/|\boldsymbol{h}_{\mathrm{so}}|$.
Note that the deformed Hamiltonian has been normalized. Accordingly, the winding number or Chern number can be defined using the normalized Hamiltonian \cite{Shiozaki2014}.

\subsection{1D topological phases characterized by the winding number}

We start with the 1D topological phases, which are the simplest cases, with the basic Hamiltonian
for the topology being two-band models. Accordingly, the Clifford matrices $\gamma_{i}$
are generated by the Pauli matrices, $\gamma_{i=0,1}\in\{\sigma_{x},\sigma_{y},\sigma_{z}\}$.
The Hamiltonian $\mathcal{H}(k)$ has chiral symmetry $\gamma=\ui\gamma_{0}\gamma_{1}$ and is characterized by the 1D winding number:
\begin{eqnarray}
\nu_{1}&\equiv&\frac{1}{4\pi\ui}\int_{{\rm BZ}}\mathrm{Tr}[\gamma\mathcal{H}\ud\mathcal{H}]\nonumber\\
&=&\frac{1}{4\pi\ui}\int_{\mathrm{BZ}}\ud k\,(h'_{0}\partial_{k}h'_{1}-h'_{1}\partial_{k}h'_{0})\mathrm{Tr}[\gamma\gamma_{0}\gamma_{1}],
\end{eqnarray}
here we have used the fact $\sum_{i=0}^{d}h_{i}^{\prime2}=1$ for
the normalized Hamiltonian~\eqref{eq:s_deformedH}. With the deformed Hamiltonian and
the trace property of Clifford matrices $\mathrm{Tr}[\gamma\gamma_{i}\gamma_{j}]=-2\ui\epsilon^{ij}$, the winding number
can be further written as
\begin{eqnarray}
\nu_{1}&=&\frac{1}{2\pi}\int_{\mathrm{BZ}}\ud\Theta_{k}\,\hat{h}_{\mathrm{so}}\nonumber\\
&=&\frac{1}{2}\sum_{\mathrm{BIS}_{j}}\bigl[\mathrm{sgn}(h_{1,R_j})-\mathrm{sgn}(h_{1,L_j})\bigr],\label{eq:nu1}
\end{eqnarray}
where the unit SO field $\hat{h}_{\mathrm{so}}=\mathrm{sgn}(h_{1})$ for 1D, with $\mathrm{sgn}(x)$ being the sign function, and the BISs become discrete points (see Fig.~5a in the main text), with the left hand and right hand points of the $j$-th BIS being denoted as $L_j$ and $R_j$, respectively. Thus the right hand side of the above formula denotes a zero-dimensional invariant ($0$-th Chern number). Note that the region ${\mathcal{V}_{\mathrm{BIS}}}$ enclosed by each
BIS refers to that with $h_{0}(k)<0$, see Fig.~5c in the main text. The integral of $\Theta_{k}$
must become $+\pi$ at the right-hand ($R_j$) and $-\pi$ at the left-hand ($L_j$) point of
the $j$-th BIS, which results in the prefactors of $\mathrm{sgn}(h_{1,R_j})$ and $\mathrm{sgn}(h_{1,L_j})$, respectively.
For the case with only a single BIS, the equation (\ref{eq:nu1}) gives the result with a simple picture that the system is topologically nontrivial
when the SO field has opposite signs at the $R$ and $L$ points of BIS, otherwise it is trivial.

As an example, we consider the 1D model of the AIII class in the main text, which has the Hamiltonian $\mathcal{H}(k)=\vec{h}(k)\cdot\vec{\sigma}$ with $h_{x}=t_{\mathrm{so}}\sin k\equiv h_{\mathrm{so}}, h_{y}=0$ and $h_{z}=m_{z}-t_{0}\cos k\equiv h_{0}$. Here $t_{0}>0$ and
$t_{\mathrm{so}}>0$ denote the nearest-neighbor spin-conserved and spin-flipped hopping coefficients, $m_{z}$ is the effective magnetization. When $|m_{z}|>t_{0}$, this model is trivial as there is no BISs. On the other hand, the BIS is determined by $k_{\mathrm{BIS}}=\pm\arccos m_{z}/t_{0}$ when $|m_{z}|<t_{0}$. It is easy to check that $h_{\mathrm{so}}>0$ for the right hand point of the BIS and $h_{\mathrm{so}}<0$ for the left hand point, which indicates the winding number $\nu_{1}=1$.

\subsection{2D topological phases characterized by the first Chern number}

For the 2D topological phases, a two-band model can be formulated with three Pauli matrices that $\gamma_{i}\in\{\sigma_{x},\sigma_{y},\sigma_{z}\}$
for $i=0,1,2$. Accordingly, the topology of the phase is characterized by the first
Chern number
\begin{equation}
\mathrm{Ch}_{1}=-\frac{\ui}{16\pi}\int_{\mathrm{BZ}}\mathrm{Tr}\bigl[\mathcal{H}(\ud\mathcal{H})^{2}\bigr].
\end{equation}
The trace is taken in the spin space. Using the relation $\mathcal{H}(\ud\mathcal{H})^{2}=\sum_{i}\sum_{j\neq k}h'_{i}\ud h'_{j}\wedge\ud h'_{k}\gamma_{i}\gamma_{j}\gamma_{k}$
and the trace property $\mathrm{Tr}[\gamma_{i}\gamma_{j}\gamma_{k}]=-2\ui\epsilon^{ijk}$,
where the Levi-Civita tensor and the wedge are both anti-symmetric, we further obtain the Chern number that
\begin{equation}
\mathrm{Ch}_{1}=-\frac{1}{4\pi}\int_{\mathrm{BZ}}\ud^{2}\mathbf{k}\cdot(h'_{0}\nabla h'_{1}\times\nabla h'_{2}-h'_{1}\nabla h'_{0}\times\nabla h'_{2}+h'_{2}\nabla h'_{0}\times\nabla h'_{1}).
\end{equation}
From the deformed Hamiltonian (\ref{eq:s_deformedH}) we obtain that
\begin{align}
h'_{0}\nabla h'_{1}\times\nabla h'_{2} & =\sin\Theta_{\mathbf{k}}\cos^{2}\Theta_{\mathbf{k}}\nabla\hat{h}_{\mathrm{so},1}\times\nabla\hat{h}_{\mathrm{so},2}\nonumber\\
&-\sin^{2}\Theta_{\mathbf{k}}\cos\Theta_{\mathbf{k}}\nabla\Theta_{\mathbf{k}}\times(\hat{h}_{\mathrm{so},1}\nabla\hat{h}_{\mathrm{so},2}-\hat{h}_{\mathrm{so},2}\nabla\hat{h}_{\mathrm{so},1}),\nonumber \\
h'_{i}\nabla h'_{0}\times\nabla h'_{j} & =\cos^{3}\Theta_{\mathbf{k}}\nabla\Theta_{\mathbf{k}}\times\hat{h}_{\mathrm{so},i}\nabla\hat{h}_{\mathrm{so},j}.\nonumber
\end{align}
With the above results the Chern number is recast into
\begin{equation}
\mathrm{Ch}_{1}=-\frac{1}{4\pi}\int_{\mathrm{BZ}}\ud^{2}\mathbf{k}\cdot\bigl[\sin\Theta_{\mathbf{k}}\cos^{2}\Theta_{\mathbf{k}}\nabla\hat{h}_{\mathrm{so},1}\times\nabla\hat{h}_{\mathrm{so},2}-\cos\Theta_{\mathbf{k}}\nabla\Theta_{\mathbf{k}}\times(\hat{h}_{\mathrm{so},1}\nabla\hat{h}_{\mathrm{so},2}-\hat{h}_{\mathrm{so},2}\nabla\hat{h}_{\mathrm{so},1})\bigr].
\end{equation}
To simplify the above formula, we can choose a special configuration of the $\Theta_{\bold k}$ function, while the final result is independent of the choice of $\Theta_{\bold k}$. For this we let the SO field ($\propto\cos\Theta_{\mathbf{k}}$) for the deformed Hamiltonian be truncated to the BISs ($\sin\Theta_{\mathbf{k}}=0$) by taking $\Theta_{\mathbf{k}}$ to
approach the step function. Then the contribution from the first term vanishes. On the other hand, $\nabla\Theta_{\mathbf{k}}$ tends
to a delta function, and the second term will contribute to the Chern
number. Therefore, the whole integral can be reduced to that for the
second term over the region $\mathcal{R}$ with non-zero $\nabla\Theta_{\mathbf{k}}$ (see Fig.~5c-d in the main text):
\begin{align}
\mathrm{Ch}_{1} & =\frac{1}{4\pi}\int_{\mathcal{R}}\ud^{2}\mathbf{k}\cdot\cos\Theta_{\mathbf{k}}\nabla\Theta_{\mathbf{k}}\times(\hat{h}_{\mathrm{so},1}\nabla\hat{h}_{\mathrm{so},2}-\hat{h}_{\mathrm{so},2}\nabla\hat{h}_{\mathrm{so},1})\nonumber \\
 & =\frac{1}{4\pi}\int_{-\pi/2}^{\pi/2}\ud\Theta_{k_{\perp}}\,\cos\Theta_{k_{\perp}}\sum_{j}\oint_{\mathrm{BIS}_{j}}\ud k_{\parallel}\,(\hat{z}\times\hat{n}_{\perp})\cdot(\hat{h}_{\mathrm{so},1}\nabla\hat{h}_{\mathrm{so},2}-\hat{h}_{\mathrm{so},2}\nabla\hat{h}_{\mathrm{so},1})\nonumber \\
 &=\sum_{j}\frac{1}{2\pi}\oint_{\mathrm{BIS}_{j}}\ud\hat{\ell}\cdot(\hat{h}_{\mathrm{so},1}\nabla\hat{h}_{\mathrm{so},2}-\hat{h}_{\mathrm{so},2}\nabla\hat{h}_{\mathrm{so},1})\nonumber\\
 &=\sum_{j}\frac{1}{2\pi}\epsilon^{mn}\oint_{\mathrm{BIS}_{j}}\ud\hat{\ell}\cdot\hat{{h}}_{\mathrm{so},m}\nabla\hat{h}_{\mathrm{so},n},\label{eq:ch2}
\end{align}
where we have decomposed the momentum $\mathbf{k}$
into $(k_{\perp},k_{\parallel})$ in the second line, with $k_{\perp}$ perpendicular
to the BIS and pointing to the side $\bar{\mathcal{V}}_{\mathrm{BIS}}$ with
$h_{0}(\mathbf{k})>0$ (see Fig.~5c in the main text), and $m,n=1,2$. From the last line of the above formula we know that the final result is independent of the function $\Theta_{\bold k}$. Equation
(\ref{eq:ch2}) shows that the Chern number in the bulk reduces to the 1D winding of the SO field $\hat{\boldsymbol{h}}_{\mathrm{so}}$
on the 1D BISs (band inversion rings). The integral $\sum_{j}\frac{1}{2\pi}\epsilon^{\alpha\beta}\oint_{\mathrm{BIS}_{j}}\ud\hat{\ell}\cdot\hat{{h}}_{\mathrm{so},\alpha}\nabla\hat{h}_{\mathrm{so},\beta}$ is performed on the 1D BISs, with the positive direction defined so that the closed integral paths denote the boundary of the vector area ${\cal V}_{\rm BIS}$. The 1D winding number then counts the total vorticity of the SO field at the degenerate points, i.e., the total topological charge of the 2D Dirac points, characterized by $\hat{\boldsymbol{h}}_{\mathrm{so}}=0$ within ${\cal V}_{\rm BIS}$. The above result implies
that, the first Chern number is simply given by the total topological charge
of the SO field in the region enclosed by BISs, namely, in the $\mathcal{V}_{\mathrm{BIS}}$
with $h_{0}(\mathbf{k})<0$.

Here we take the Chern insulator in the main text as an example, whose Hamiltonian can be written as $\mathcal{H}(\mathbf{k})=\vec{h}(\mathbf{k})\cdot\vec{\sigma}$ with $\vec{h}(\mathbf{k})=(t_{\mathrm{so}}\sin k_{x},t_{\mathrm{so}}\sin k_{y}, m_{z}-t_{0}\cos k_{x}-t_{0}\cos k_{y})$. Without loss of the generality, we can decompose the vector field $\vec{h}$ into $h_{0}\equiv h_{z}$ and $\boldsymbol{h}_\mathrm{{so}}\equiv (h_{y},h_{x})$. When $m_{z}>2$, there is no BISs and the system is trivial. When $0<m_{z}<2$, a topological charge $-1$ located at $\mathbf{k}=0$ in ${\cal V}_{\rm BIS}$ is enclosed by the BIS, and gives the Chern number $\mathrm{Ch}_{1}=-1$, see Fig.~6a in the main text. Furthermore, another two topological charges $+1$ at $(0,\pi)$ and $(\pi,0)$ are also included when $-2<m_{z}<0$, which renders the Chern number being changed to $1+1-1=+1$. Finally, when $m_{z}<-2$, the region $\mathcal{V}_{\mathrm{BIS}}$ equals the whole BZ, and all the four Dirac points $\{(0,0),(\pi,\pi),(0,\pi),(\pi,0)\}$ should be considered and contribute a trivial total topological charge ($1+1-1-1=0$).

On the other hand, one can also choose that $h_{0}\equiv t_{\mathrm{so}}\sin k_{x}$, in which case $\mathcal{V}_{\mathrm{BIS}}$
with $h_{0}(\mathbf{k})<0$ corresponds to the region with $-\pi<k_{x}<0$, and then
the SO field is $\boldsymbol{h}_{\mathrm{so}}\equiv(h_z,h_y)$.
A pair of topological charges with opposite signs, located at $\mathbf{k}=\left(\pm\arccos[\left(m_{z}+1\right)/t_{0}],\pi\right)$ for $-2<m_z<0$
and at $\bold k=\left(\pm\arccos[(m_{z}-1)/t_{0}],0\right)$ for $0<m_z<2$, will be created from vacuum,
moves and finally annihilates, as $m_{z}$ varies from $-\infty$ to
$+\infty$. Only one of the two charges is located in $\mathcal{V}_{\mathrm{BIS}}$ and enclosed by the BIS, giving the Chern number $C_1=+1$ (for $-2<m_z<0$) or $C_1=-1$ (for $0<m_z<2$) (see Fig.~6b in the main text). The results are clearly consistent with those given above by choosing $h_{0}=h_z(\bold k)$.

\subsection{3D topological phases characterized by the winding number}

For 3D topological phases, the minimal Hamiltonian involves four
bands, and the Clifford matrices take the Dirac forms. The Hamiltonian
has chiral symmetry $\gamma=-\prod_{i=0}^{3}\gamma_{i}$ and is characterized
by the 3D winding number
\begin{equation}
\nu_{3}=\frac{1}{48\pi^{2}}\int_{\mathrm{BZ}}\mathrm{Tr}\bigl[\gamma\mathcal{H}(\ud\mathcal{H})^{3}\bigr].
\end{equation}
Similar to the 2D case, the relation $\mathcal{H}(\ud\mathcal{H})^{3}=\sum_{i}\sum_{j\neq s\neq t}h'_{i}\ud h'_{j}\wedge\ud h'_{s}\wedge\ud h'_{t}$
and the trace gives $\mathrm{Tr}[\gamma\gamma_{i}\gamma_{j}\gamma_{s}\gamma_{t}]=-4\epsilon^{ijst}$, with which
the winding number takes the following explicit form
\begin{equation}
\nu_{3}=-\frac{1}{2\pi^{2}}\int_{\mathrm{BZ}}\ud^{3}\mathbf{k}\,\Bigl\{ h'_{0}\nabla h'_{1}\cdot(\nabla h'_{2}\times\nabla h'_{3})-\nabla h'_{0}\cdot\bigl[h'_{1}\nabla h'_{2}\times\nabla h'_{3}-h'_{2}\nabla h'_{1}\times\nabla h'_{3}+h'_{3}\nabla h'_{1}\times\nabla h'_{2}\bigr]\Bigr\}.
\end{equation}
From the deformed Hamiltonian (\ref{eq:s_deformedH}), the integrand in the winding number can be simplified by
\begin{align}
h'_{0}\nabla h'_{1}\cdot(\nabla h'_{2}\times\nabla h'_{3})= & \sin\Theta_{\mathbf{k}}\cos^{3}\Theta_{\mathbf{k}}\nabla\hat{h}_{\mathrm{so},1}\cdot(\nabla\hat{h}_{\mathrm{so},2}\times\nabla\hat{h}_{\mathrm{so},3})\nonumber \\
 & -\sin^{2}\Theta_{\mathbf{k}}\cos^{2}\Theta_{\mathbf{k}}\nabla\Theta_{\mathbf{k}}\cdot(\hat{h}_{\mathrm{so},1}\nabla\hat{h}_{\mathrm{so},2}\times\nabla\hat{h}_{\mathrm{so},3}\nonumber\\
 &-\hat{h}_{\mathrm{so},2}\nabla\hat{h}_{\mathrm{so},1}\times\nabla\hat{h}_{\mathrm{so},3}+\hat{h}_{\mathrm{so},3}\nabla\hat{h}_{\mathrm{so},1}\times\nabla\hat{h}_{\mathrm{so},2}),\nonumber \\
h'_{i}\nabla h'_{0}\cdot(\nabla h'_{j}\times\nabla h'_{k})= & \cos^{4}\Theta_{\mathbf{k}}\nabla\Theta_{\mathbf{k}}\cdot(\hat{h}_{\mathrm{so},i}\nabla\hat{h}_{\mathrm{so},j}\times\nabla\hat{h}_{\mathrm{so},k}).\nonumber
\end{align}
Similar to the derivation done for the 2D phase, we take a special configuration that $\Theta_{\mathbf{k}}$ approaches the step function. The computation of the winding number finally reduces to the integral on the BISs, given by
\begin{align}
\nu_{3} & =\frac{1}{2\pi^{2}}\int_{\mathcal{R}}\ud^{3}\mathbf{k}\,\cos^{2}\Theta_{\mathbf{k}}\nabla\Theta_{\mathbf{k}}\cdot\bigl[\hat{h}_{\mathrm{so},1}\nabla\hat{h}_{\mathrm{so},2}\times\nabla\hat{h}_{\mathrm{so},3}-\hat{h}_{\mathrm{so},2}\nabla\hat{h}_{\mathrm{so},1}\times\nabla\hat{h}_{\mathrm{so},3}+\hat{h}_{\mathrm{so},3}\nabla\hat{h}_{\mathrm{so},1}\times\nabla\hat{h}_{\mathrm{so},2}\bigr]\nonumber \\
 & =\frac{1}{2\pi^{2}}\int_{-\pi/2}^{\pi/2}\ud\Theta_{k_{\perp}}\,\cos^{2}\Theta_{k_{\perp}}\sum_{j}\int_{\mathrm{BIS}_{j}}\ud^{2}\boldsymbol{k}_{\parallel}\cdot\bigl[\hat{h}_{\mathrm{so},1}\nabla\hat{h}_{\mathrm{so},2}\times\nabla\hat{h}_{\mathrm{so},3}-\hat{h}_{\mathrm{so},2}\nabla\hat{h}_{\mathrm{so},1}\times\nabla\hat{h}_{\mathrm{so},3}+\hat{h}_{\mathrm{so},3}\nabla\hat{h}_{\mathrm{so},1}\times\nabla\hat{h}_{\mathrm{so},2}\bigr]\nonumber \\
 & =\sum_{j}\frac{1}{4\pi}\int_{\mathrm{BIS}_{j}}\ud^{2}\boldsymbol{k}_{\parallel}\cdot\bigl[\hat{h}_{\mathrm{so},1}\nabla\hat{h}_{\mathrm{so},2}\times\nabla\hat{h}_{\mathrm{so},3}-\hat{h}_{\mathrm{so},2}\nabla\hat{h}_{\mathrm{so},1}\times\nabla\hat{h}_{\mathrm{so},3}+\hat{h}_{\mathrm{so},3}\nabla\hat{h}_{\mathrm{so},1}\times\nabla\hat{h}_{\mathrm{so},2}\bigr]\nonumber \\
 & =\sum_{j}\frac{1}{8\pi}\epsilon^{mnl}\int_{\mathrm{BIS}_{j}}\ud^{2}\boldsymbol{k}\cdot\hat{{h}}_{\mathrm{so},m}(\nabla\hat{{h}}_{\mathrm{so}.n} \times\nabla\hat{{h}}_{\mathrm{so},l}).
\end{align}
The last line of the above equation denotes the coverage of the SO field $\hat{\boldsymbol{h}}_{\mathrm{so}}(\bold k)$ over the 2D spherical surface $S^2$ when $\bold k$ runs over the BISs, which implies that the 3D winding number of the bulk is mapped to the 2D Chern number defined on the 2D BISs.
As is known that the integral $\frac{1}{8\pi}\epsilon^{mnl}\int_{\mathrm{BIS}_{j}}\ud^{2}\boldsymbol{k}\cdot\hat{{h}}_{\mathrm{so},m}(\nabla\hat{{h}}_{\mathrm{so}.n} \times\nabla\hat{{h}}_{\mathrm{so},l})$
counts the topological charge of the Weyl points at $\boldsymbol{h}_{\mathrm{so}}(\mathbf{k})=0$ of the SO field in the vector area ${\cal V}_{\rm BIS}$ enclosed by the BISs.

We also give an example for the 3D topological phases. Consider the
B phase of superfluid $^{3}$He \cite{Chiu2016,Volovik2003}, which can be treated as a 3D topological
superconducting phase in class DIII. The Hamiltonian reads $\mathcal{H}(\mathbf{k})=h_{0}(\mathbf{k})\sigma_{z}\otimes\mathbbm{1}+h_{1}(\mathbf{k})\sigma_{x}\otimes\sigma_{z}+h_{2}(\mathbf{k})\sigma_{y}\otimes\mathbbm{1}+h_{3}(\mathbf{k})\sigma_{x}\otimes\sigma_{x}$
with $\vec{h}(\mathbf{k})=(\mathbf{k}^{2}/2m-\mu,k_{x},-k_{y},-k_{z})$.
It is easy to see that the system is trivial for $\mu<0$. For the case with $\mu>0$, a single Weyl
point with $\boldsymbol{h}_{\mathrm{so}}=(h_1,h_2,h_3)=0$ at $\bold k=0$ and of charge $+1$ is enclosed by the BIS, which renders the system being topologically nontrivial with winding number $\nu_{3}=+1$.

With the above detailed proof for the 1D, 2D and 3D topological phases, the generic proof for bulk-surface duality in the Appendix can be better understood.

\section{Characterizing quantum anomalous Hall insulators}

\begin{figure}
\includegraphics[width=0.618\textwidth]{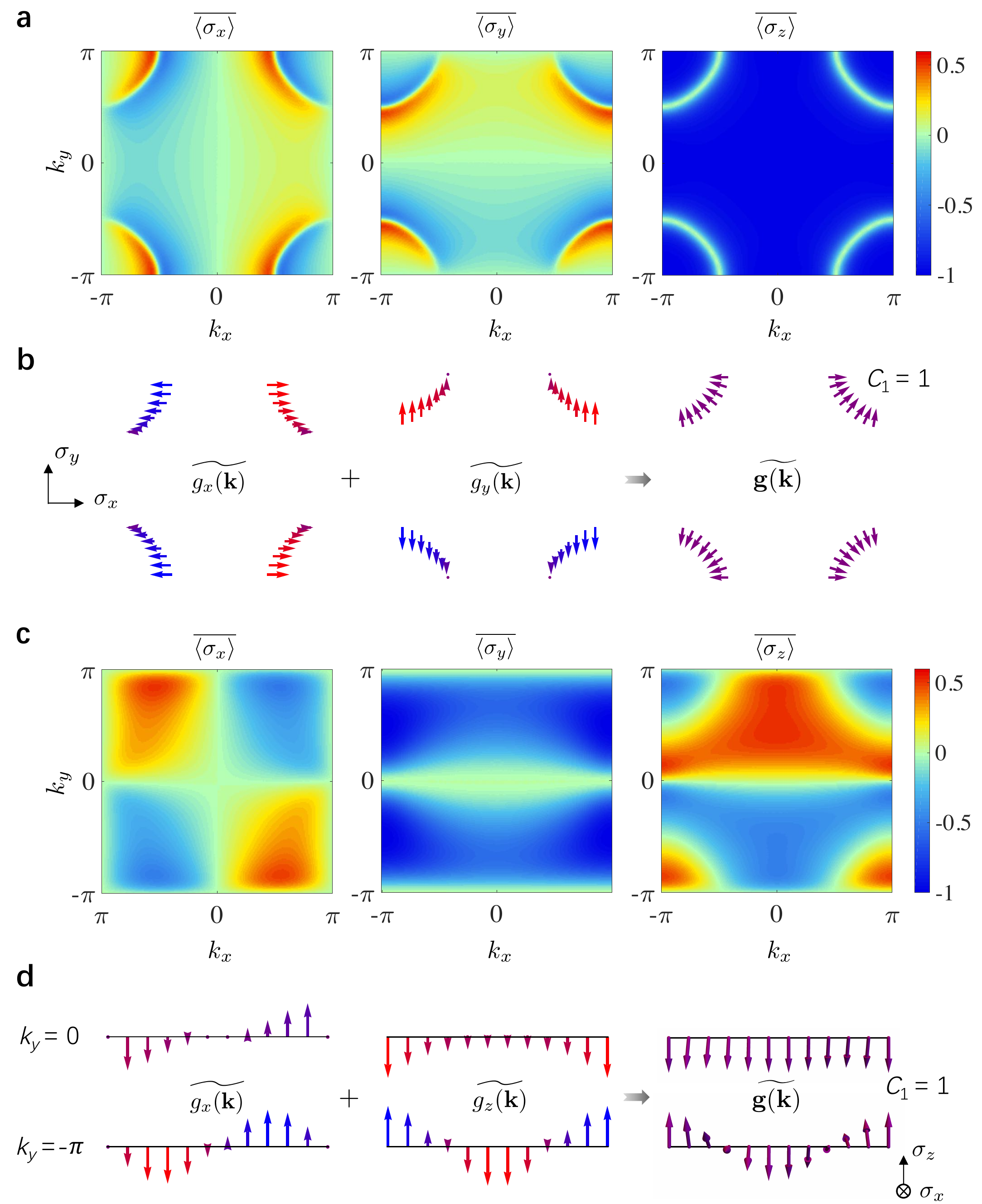}
\caption{\textsf{\textbf{Quenching to the topological phase with $C_1=1$}}.
\textsf{\textbf{a-b}}, Quenching the $h_z$ axis.  Time-averaged spin textures are measured after a sudden quench from $m_z=8t_0$ to $-t_0$ ({\sf\textbf{a}}), with BISs being characterized by $\overline{\langle\vec\sigma\rangle}=0$ (the green ring centered at $(\pi,\pi)$).
The $\widetilde{{\bold g}(\mathbf{k})}$ field is obtained by combining the derivatives $\partial_{k_\perp}\!\overline{\langle\sigma_{x,y}\rangle}$ ({\sf\textbf{b}}), and gives the Chern number $C_1=1$. Here $t_{\mathrm{so}}^{x,y}$ both take $0.2t_{0}$. \textsf{\textbf{c-d}}, Quenching the $h_y$ axis. Time-averaged spin textures ({\sf\textbf{c}}) and the dynamical field $\widetilde{{\bold g}(\mathbf{k})}$ on BISs ({\sf\textbf{d}})
are shown after a quench from $m_y=50t_0$ to $0$ with $m_z=-t_0$. The BISs determined by $\overline{\langle\vec\sigma\rangle}=0$
are identified to $k_y=0,-\pi$, and the non-zero winding number of the dynamical spin-texture field along
$k_y=-\pi$ characterizes the Chern number $C_1=1$. Here we set $t_{\rm so}^x=t_0$ and $t_{\rm so}^y=2t_0$.
}\label{fig:figs1}
\end{figure}

The 2D quantum anomalous Hall (QAH) model generally reads ${\cal H}({\bf k})={\vec h}({\bf k})\cdot{\vec \sigma}$ with ${\vec h}({\bf k})=(m_x+t_{\rm so}^x\sin k_x, m_y+t_{\rm so}^y\sin k_y, m_z-t_0\cos k_x-t_0\cos k_y)$, which has been realized in Ref.~\cite{Wu2016,Sun2017}.
Here we identify the topological phases by $m_z$ ($m_x=m_y=0$): (i) $0<m_z<2t_0$ with the Chern number $C_1=-1$; (ii) $-2t_0<m_z<0$ with $C_1=1$; (iii) being trivial for $|m_{z}|>2t_{0}$.
The flexibility in the decomposition of the vector ${\vec h}({\bf k})$ allows us to consider different quench ways to detect the topology.
As discussed in the main text, we can take $h_0\equiv h_z, \bold h_{\rm so}\equiv(h_x,h_y)$, and the quench is performed by varying $m_z$, which corresponds to quenching the
$h_{z}$ axis. Alternatively, we can also take $h_0\equiv h_y$ and quench the $h_{y}$ axis by ramping $m_y$ quickly from $|m_y|\gg t_0$ to $0$ while fixing $m_z$.
These two quench ways leads to distinct observations, but can be both used to distinguish different topological phases.
The only thing to be noted is that the ratios $t_{\rm so}^{x,y}/t_0$ should be properly adjusted to clarify the observation in the two quench ways.

When quenching the $h_z$ axis, we set $t_{\rm so}^{x}=t_{\rm so}^{y}=0.2t_0$, and suddenly change $m_z$ from the deep trivial phase $|m_z|\gg2t_0$
to one of the two topological phases.
In Fig.~2a-b of the main text, we have shown the results of quenching to the phase (i). Here we provide the results of the phase (ii) in Fig.~\ref{fig:figs1}a-b.  As discussed in the former
sections, the vanishing time-averaged spin texture $\overline{\langle\vec\sigma\rangle}=0$ characterizes the BISs, and the dynamical spin-texture field, defined as $\widetilde{{\bold g}(\mathbf{k})}=-\partial_{k_\perp}\overline{\langle\vec\sigma\rangle}$ across the BISs, signifies the SO field $\hat{{\boldsymbol h}}_{\mathrm{so}}(\mathbf{k})$
and can be used to classify the topology.
Compared to Fig.~2a-b in the main text, the BIS in Fig.~\ref{fig:figs1}a becomes a ring centered at $\mathbf{k}=(\pi,\pi)$ rather than at $(0,0)$. Here the region enclosed by the BIS, i.e. ${\cal V}_{\rm BIS}$, contains three topological charges of the SO field: one charge $-1$ located at $(0,0)$ and other two charges $+1$ at $(0,\pi)$ and $(\pi,0)$, which indicates the Chern number $C_{1}=1$ and can be reflected by the winding of the spin-texture field $\widetilde{{\bold g}(\mathbf{k})}$ along the BIS.

When quenching the $h_y$ axis, we set $t_{\rm so}^{x}=t_0$ and $t_{\rm so}^{y}=2t_0$.
In this quench way, the BISs are the same for different topological regimes, and are identified to two lines $k_y=-\pi$ and $k_y=0$.
However the dynamical spin-texture field on the BISs should be different, which classifies the different topology, as shown in Fig.~2c-d for $C_{1}=-1$ and in Fig.~\ref{fig:figs1}c-d for $C_{1}=1$. In Fig.~2c-d, the emergent dynamical spin-texture field $\widetilde{\bold{g}(\mathbf{k})}$ winds clockwise along $k_{y}=0$ in the $\sigma_{x}$-$\sigma_{z}$ plane while the winding along $k_y=-\pi$ is zero. The situation becomes reverse in Fig.~\ref{fig:figs1}c-d, where the spin-texture field $\widetilde{\bold{g}(\mathbf{k})}$ winds anti-clockwise along $k_{y}=-\pi$ and becomes trivial along $k_{y}=0$, which characterizes the Chern number $C_{1}=1$.

\section{2D topologically trivial phase}

\begin{figure}
\includegraphics[width=0.618\textwidth]{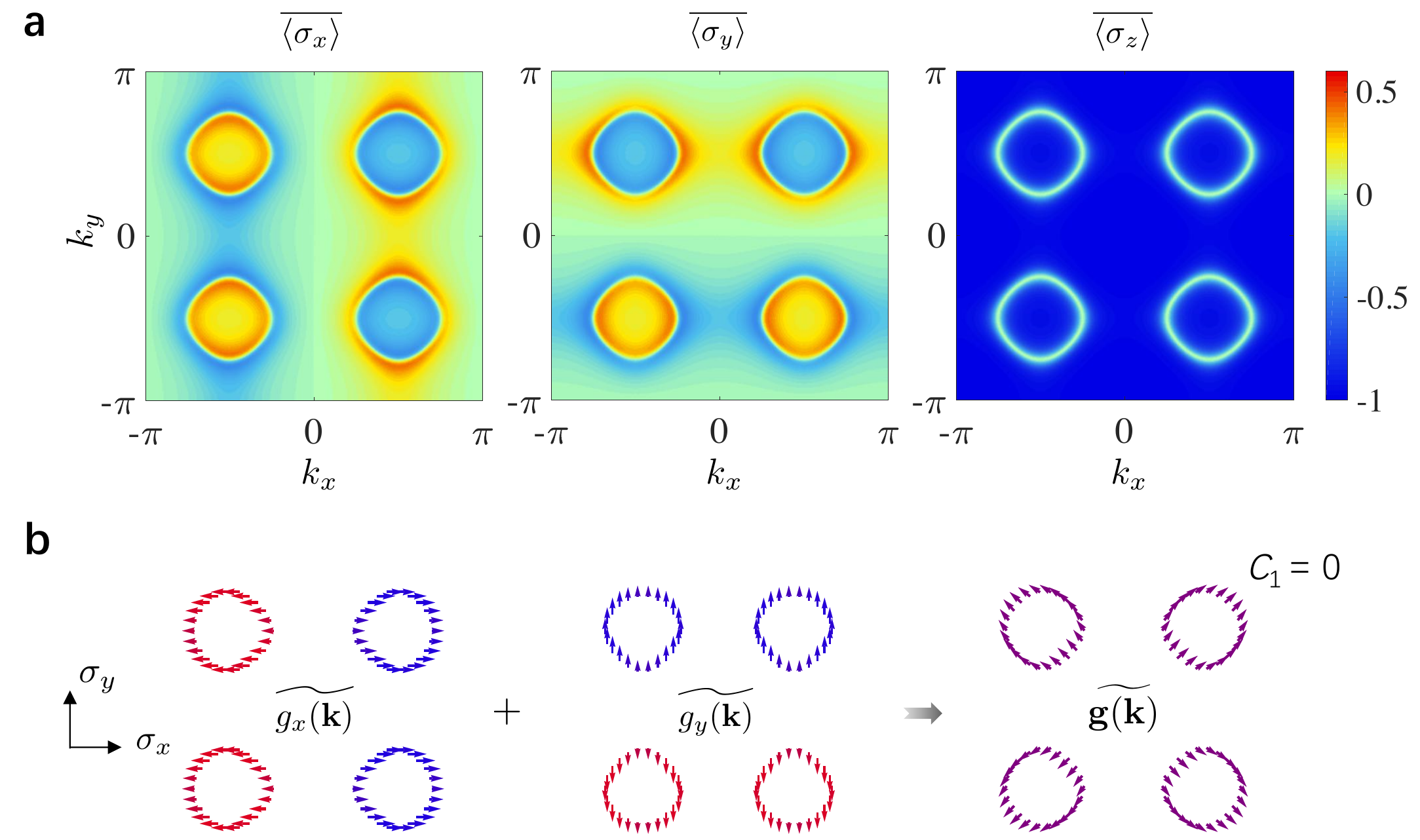}
\caption{\textsf{\textbf{2D topologically trivial phase}}.
\textsf{\textbf{a-b}}, Dynamical results of the model ${\cal H}_{\rm tr}$.
Time-averaged spin textures ({\bf{\textsf a}}) show four rings, and all contribute no winding ({\bf{\textsf b}}), charactering the trivial topology.
The quench is taken from $m_z=8t_0$ to $-t_0$ with $t_{\rm so}=0.2t_0$.
}\label{figs2}
\end{figure}

It seems that the post-quench phase always has non-trivial topology as long as the BIS is observed. However, that is not true.
As a counterexample, we consider the topologically trivial model ${\cal H}_{\rm tr}={\vec h}({\bf k})\cdot{\vec\sigma}$, where the vector field reads ${\vec h}({\bf k})=(t_{\rm so}\sin k_x,t_{\rm so}\sin k_y,m_z-t_0\cos 2k_x-t_0\cos 2k_y)$.
After the sudden quench from $m_z\gg 2t_0$ to $|m_z|<2t_0$, the time-averaged spin textures also exhibit four rings (see Fig.~\ref{figs2}a), but
the dynamical spin-texture field contribute no winding, charactering the trivial topology (Fig.~\ref{figs2}b).

\section{Detecting topology via dissipative dynamics}

\begin{figure}
\includegraphics[width=0.65\textwidth]{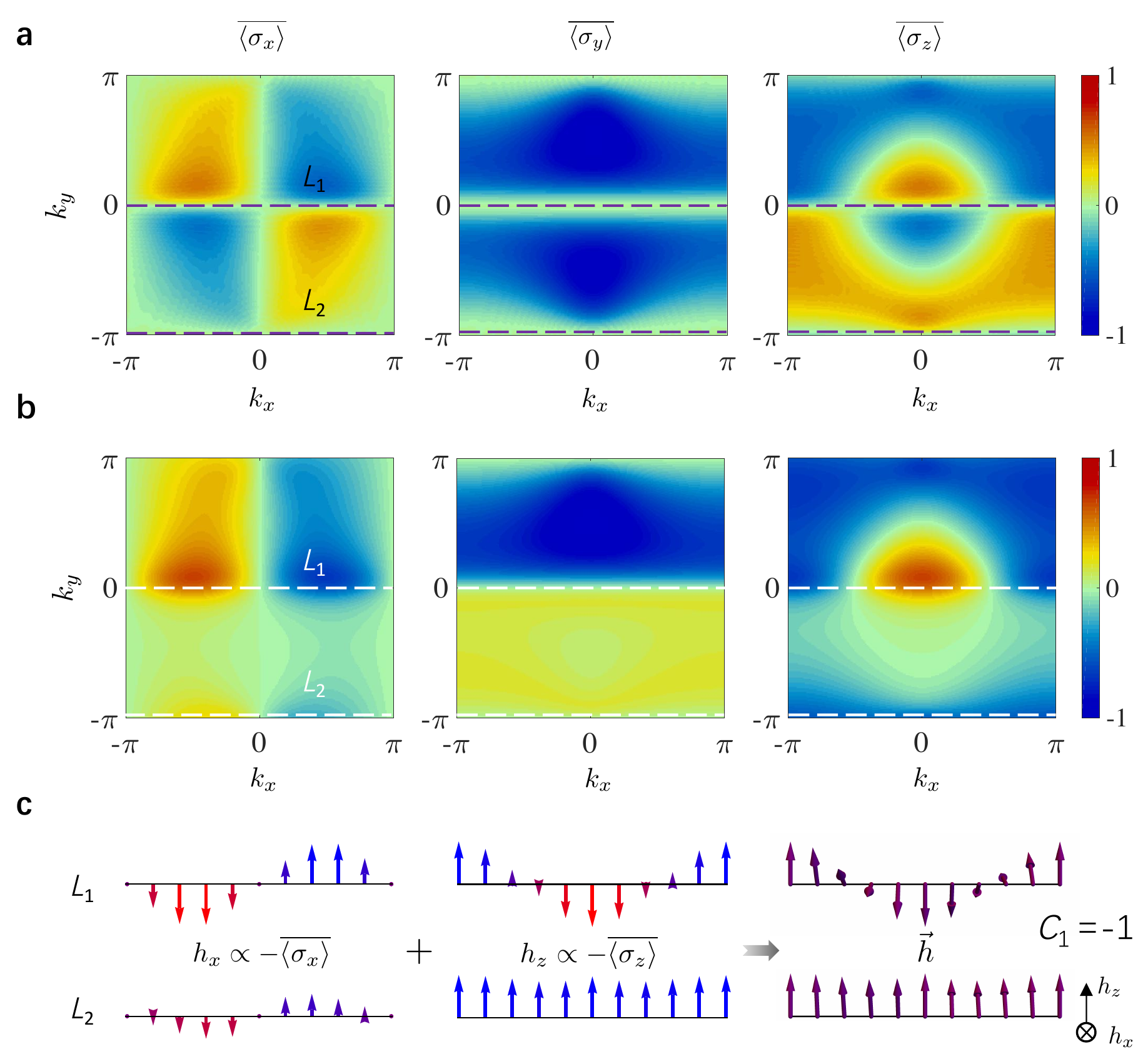}
\caption{\textsf{\textbf{Classifying topology via dissipative dynamics after quench in $h_y$ axis}}.
\textsf{\textbf{a}}, Spin textures of the 2D model averaged over a short period of $25/t_0$, which can be used to identify the BIS
by $\overline{\langle\vec\sigma\rangle}=0$. Here the two lines $L_{1,2}$ (dashed) are approximate to the BISs.
\textsf{\textbf{b}}, Time-averaged spin textures over a long
period of $600/t_0$, which directly indicate the SO field on the BISs (see {\bf{\textsf c}}).
\textsf{\textbf{c}}, The direction of the SO field is reflected by $h_{x,z}\propto -\overline{\langle\sigma_{x,z}\rangle}$.
Thus the winding of the SO field along the BISs can be read out from the spin polarizations (in {\bf{\textsf b}}) on the two lines $L_{1,2}$.
The non-zero winding along $L_1$ characterizes the Chern number $C_1=-1$.
Here the dissipative rate is $\eta=0.003$. Other quench parameters are taken the same as in Fig.~2c of the main text.
}\label{fig:figs3}
\end{figure}

The dynamical classification can be generalized to the dissipative quantum dynamics at finite temperature. The spin dynamics can be described by the so-called Lindblad mater equation~\cite{Lindblad,Hu2016}. As shown in the main text, after a such quench, the dissipation
drives the occupation of Bloch states to approach the equilibrium result,
and the spin polarization $\langle\vec\gamma\rangle$ approaches to be polarized in the reverse direction of the vector ${\vec h}({\bf k})$ after a long-time evolution.
Therefore, the dissipation can be exploited as
an alternative tool to detect the topology.

In Fig.~4 of the main text, we take the 2D model as an example, and illustrate the detecting method by quenching the $h_z$ axis.
Here we consider another case of quenching the $h_y$ axis (in contrast to the results shown in Fig.~2c-d).
In Fig.~\ref{fig:figs3}a, we present short-time-averaged spin textures, where two lines $L_{1,2}$ that satisfy $\overline{\langle\vec\sigma\rangle}=0$  can be treated as the BISs approximately. After taking long-time averages (Fig.~\ref{fig:figs3}b), we obtain non-zero spin polarizations $\overline{\langle\sigma_{x,z}\rangle}$ on the BISs,
which directly reveal the direction of the vector field ${\vec h}({\bf k})={\bf h}_{\rm so}({\bf k})$ by $h_{x,z}\propto -\overline{\langle\sigma_{x,z}\rangle}$. The winding of the SO field on the BISs is readily obtained via the spin polarizations on the two lines $L_{1,2}$ (Fig.~\ref{fig:figs3}c), and the non-zero winding on $L_1$ or $L_2$ then characterize the
topological number of the post-quench phase.


\begin{thebibliography}{99}
\bibitem{Klitzing1980} K. v. Klitzing, G. Dorda, M. Pepper, New Method for High-Accuracy Determination of the Fine-Structure Constant Based on Quantized Hall Resistance. {\em Phys. Rev. Lett.} \textbf{45}, 494 (1980).
\bibitem{Tsui1982} D. C. Tsui, H. L. Stormer, A. C. Gossard, Two-dimensional magnetotransport in the extreme quantum limit. {\em Phys. Rev. Lett.} {\bf 48}, 1559 (1982).
\bibitem{Thouless1982} D. J. Thouless {\em et al.}, Quantized Hall Conductance in a Two-Dimensional Periodic Potential. {\em Phys. Rev. Lett.} \textbf{49}, 405 (1982).
\bibitem{Wen1990} X.-G. Wen, Topological orders in rigid states. {\em Int. J. of Mod. Phys. B} {\bf 4}, 239-271 (1990).
\bibitem{Landau1999} L. D. Landau, E. M. Lifshitz, and M. Pitaevskii, 1999, Statistical
Physics (Butterworth-Heinemann, New York).

\bibitem{Hasan2010} M. Z. Hasan and C. L. Kane, Topological insulators. {\em Rev. Mod. Phys.} \textbf{82}, 3045 (2010).
\bibitem{Qi2011} X. L. Qi and S. C. Zhang, Topological insulators and superconductors. {\em Rev. Mod. Phys.} \textbf{83}, 1057 (2011).
\bibitem{Konig2007} M. K\"{o}nig {\em et al.}, Quantum Spin Hall Insulator State in HgTe Quantum Wells. {\em Science} {\bf 318}, 766 (2007).
\bibitem{Hsieh2008} D. Hsieh {\em et al.}, A topological Dirac insulator in a quantum spin Hall phase (experimental realization of a 3D Topological Insulator). {\em Nature} {\bf 452}, 970 (2008).
\bibitem{Xia2009} Y. Xia {\em et al.}, Observation of a large-gap topological-insulator class with a single Dirac cone on the surface. {\em Nat. Phys.} {\bf 5}, 398 (2009).
\bibitem{Chang2013} C.-Z. Chang {\em et al.}, Experimental Observation of the Quantum Anomalous Hall Effect in a Magnetic Topological Insulator.  {\em Science} {\bf 340}, 167 (2013).
\bibitem{Xu2015} S.-Y. Xu {\em et al.}, Discovery of a Weyl Fermion Semimetal and Topological Fermi Arcs. {\em Science} \textbf{349}, 613 (2015).
\bibitem{Lv2015} B. Q. Lv {\em et al.}, Experimental discovery of weyl semimetal TaAs. {\em Phys. Rev. X.} \textbf{5}, 31013 (2015).

\bibitem{Benevig2017} B. Bradlyn {\em et al.}, 
Topological quantum chemistry. {\em Nature} {\bf 547}, 298 (2017).

\bibitem{Kitaev2001} A. Kitaev, Unpaired Majorana fermions in quantum wires. {\em Physics-Uspekhi} \textbf{44}, 131 (2001).
\bibitem{Reed-Green2000} N. Read and D. Green, Paired states of fermions in two dimensions with breaking of parity and time-
reversal symmetries and the fractional quantum Hall effect. {\em Phys. Rev. B} {\bf 61}, 10267 (2000).

\bibitem{Alicea2012} J. Alicea, New directions in the pursuit of Majorana fermions in solid state systems. {\em Rep. Prog. Phys.} \textbf{75}, 76501 (2012).

\bibitem{Elliott2014} S. R. Elliott, M. Franz, Colloquium: Majorana Fermions in nuclear, particle and solid-state physics. {\em Rev. Mod. Phys.} \textbf{87}, 137 (2014).

\bibitem{Sato2017} M. Sato and Y. Ando, Topological superconductors: a
review. {\em Rep. Prog. Phys.} {\bf 80}, 076501 (2017).

\bibitem{Chen2012} X. Chen {\em et al.}, Symmetry-Protected Topological Orders in Interacting Bosonic Systems. {\em Science} \textbf{338}, 1604 (2012).
\bibitem{Chen2013} X. Chen {\em et al.}, Symmetry protected topological orders and the group cohomology of their symmetry group. {\em Phys. Rev. B} \textbf{87}, 155114 (2013).

\bibitem{quench1}  {L. D'Alessio and M. Rigol, Dynamical preparation of Floquet Chern insulators. Nat. Commun. {\bf 6}, 8336 (2015).}

\bibitem{quench2}  {M. D. Caio, N. R. Cooper, and M. J. Bhaseen, Quantum Quenches in Chern Insulators. Phys. Rev. Lett. {\bf 115}, 236403 (2015).}

\bibitem{quench3}  {Y. Hu, P. Zoller, and J. C. Budich, Dynamical Buildup of a Quantized Hall Response from Nontopological States. Phys. Rev. Lett. {\bf 117}, 126803 (2016).}

\bibitem{quench4}  {F. N. \"{U}nal, E. J. Mueller, and M. \"{O}. Oktel, Nonequilibrium fractional Hall response after a topological quench. Phys. Rev. A {\bf 94}, 053604 (2016).}

\bibitem{quench5}  {J. H. Wilson, J. C.W. Song, and G. Refael, Remnant Geometric Hall Response in a Quantum Quench. Phys. Rev. Lett. {\bf 117}, 235302 (2016).}


\bibitem{Su1980} W. P. Su {\em et al.}, Soliton excitations in polyacetylene. {\em Phys. Rev. B} \textbf{22}, 2099 (1980).

\bibitem{Atala2012} M. Atala {\em et al.}, Direct Measurement of the Zak phase in Topological Bloch Bands. {\em Nat. Phys.} \textbf{9}, 795 (2012).

\bibitem{Liu2013} X.-J. Liu, Z.-X. Liu, and M. Cheng, Manipulating Topological Edge Spins in a One-Dimensional Optical Lattice. {\em Phys. Rev. Lett.} {\bf 110}, 076401 (2013).

\bibitem{Song2017} B. Song, L. Zhang, C. He, T. F. J. Poon, E. Hajiyev, S. Zhang, X.-J. Liu, and G.-B. Jo,
Observation of symmetry-protected topological band with ultracold fermions. {\em Science Advances} {\bf 4},
eaao4748 (2018).

\bibitem{Jotzu2014} G. Jotzu {\em et al.}, Experimental realization of the topological Haldane model with ultracold fermions. {\em Nature} {\bf 515}, 237-240 (2014).
\bibitem{Aidelsburger2015} M. Aidelsburger {\em et al.}, Measuring the Chern number of Hofstadter bands with ultracold bosonic atoms. {\em Nat. Phys.} {\bf 11}, 162-166 (2015).

\bibitem{Wu2016} Z. Wu, L. Zhang, W. Sun, X.-T. Xu, B.-Z. Wang, S.-C. Ji, Y. Deng, S. Chen, X.-J. Liu, and J.-W. Pan, Realization of two-dimensional spin-orbit coupling for Bose-Einstein condensates.  {\em Science} {\bf 354}, 82 (2016).

\bibitem{Sun2017} W. Sun {\em et al.}, Long-lived 2D Spin-Orbit coupled Topological Bose Gas. arXiv:1710.00717.

\bibitem{Tarnowski2017} M. Tarnowski {\em et al.}, Characterizing topology by dynamics: Chern number from linking number. arXiv: 1709.01046.

\bibitem{Wang2017} C. Wang {\em et al.}, Scheme to Measure the Topological Number of a Chern Insulator from Quench Dynamics.  {\em Phys. Rev. Lett.} {\bf 118}, 185701 (2017).

\bibitem{Sunwei2018} W. Sun {\em et al.,} Uncovering topology by quenching dynamics. arXiv:1804.08226v2.

\bibitem{AZ1997} A. Altland and Martin R. Zirnbauer, Nonstandard symmetry classes in mesoscopic normal-superconducting hybrid structures. {\em Phys. Rev. B} {\bf 55}, 1142 (1997).
\bibitem{Schnyder2008} A. P. Schnyder {\em et al.}, Classification of topological insulators and superconductors in three spatial dimensions. {\em Phys. Rev. B} \textbf{78}, 195125 (2008).
\bibitem{Kitaev2009} A. Kitaev, Periodic table for topological insulators and superconductors. {\em AIP Conference Proceedings} \textbf{1134}, 22 (2009).
\bibitem{Chiu2016} C.-K. Chiu {\em et al.}, Classification of topological quantum matter with symmetries. {\em Rev. Mod. Phys.} \textbf{88}, 035005 (2016).
\bibitem{Morimoto2013} T. Morimoto {\em et al.}, Topological classification with additional symmetries from Clifford algebras. {\em Phys. Rev. B} \textbf{88}, 125129 (2013).
\bibitem{Chiu2013} C.-K. Chiu {\em et al.}, Classification of topological insulators and superconductors in the presence of reflection symmetry. {\em Phys. Rev. B} \textbf{88}, 075142 (2013).

\bibitem{SI} See Supplementary Material for more details.

\bibitem{Haldane1988} F. D. M. Haldane, Model for a Quantum Hall Effect without Landau Levels: Condensed-Matter Realization of the ``Parity Anomaly''. {\em Phys. Rev. Lett.} \textbf{61}, 2015 (1988).
\bibitem{Volovik2003} G. Volovik, The universe in a helium droplet (Clarendon Press, 2003).

\bibitem{Zhang2001} S. C. Zhang and J. Hu, A four-dimensional generalization of the quantum Hall effect. {\em Science}, {\bf 294}, 823-828 (2001).


\bibitem{Liu2014} X.-J. Liu, K. T. Law, and T. K. Ng, Realization of 2D
Spin-Orbit Interaction and Exotic Topological Orders in Cold Atoms. {\em Phys. Rev. Lett.} {\bf 112}, 086401 (2014).

\bibitem{Wang2018} B.-Z. Wang, Y.-H. Lu, W. Sun, S. Chen, Y. Deng, X.-J. Liu, Dirac-, Rashba-, and Weyl-type spin-orbit couplings: Toward experimental realization in ultracold atoms. {\it Phys. Rev. A} {bf 97}, 011605(R) (2018).

\bibitem{Lindblad} G. Lindblad, On the generators of quantum dynamical semigroups. {\em Commun. Math. Phys.}
{\bf 48}, 119-130 (1976).


\bibitem{Shiozaki2014} K. Shiozaki {\em et al.}, Topology of crystalline insulators and superconductors. {\em Phys. Rev. B} \textbf{90}, 165114 (2014).

\bibitem{Slager2013} R. -J. Slager 
{\em et al.}, The space group classification of topological
band-insulators. {\em Nat. Phys.} {\bf 9}, 98 (2013).

\bibitem{Chan2017-2} C. Chan, L. Zhang, T.-F. J. Poon, Y.-P. He, Y.-Q. Wang, and X.-J. Liu, Generic Theory for Majorana Zero Modes in 2D Superconductors. {\em Phys. Rev. Lett.} 119, 047001 (2017).

\end{thebibliography}

\begin{thebibliography}{99}
\bibitem[S1]{Shiozaki2014} K. Shiozaki {\em et al.}, Topology of crystalline insulators and superconductors. {\em Phys. Rev. B} \textbf{90}, 165114 (2014).

\bibitem[S2]{Chiu2016} C.-K. Chiu {\em et al.}, Classification of topological quantum matter with symmetries. {\em Rev. Mod. Phys.} \textbf{88}, 035005 (2016).

\bibitem[S3]{Volovik2003} G. Volovik, The universe in a helium droplet (Clarendon Press, 2003).

\bibitem[S4]{Wu2016} Z. Wu, L. Zhang, W. Sun, X.-T. Xu, B.-Z. Wang, S.-C. Ji, Y. Deng, S. Chen, X.-J. Liu, and J.-W. Pan, Realization of two-dimensional spin-orbit coupling for Bose-Einstein condensates.  {\em Science} {\bf 354}, 82 (2016).

\bibitem[S5]{Sun2017} W. Sun {\em et al.}, Long-lived 2D Spin-Orbit coupled Topological Bose Gas. arXiv:1710.00717.

\bibitem[S6]{Lindblad} G. Lindblad, On the generators of quantum dynamical semigroups. {\em Commun. Math. Phys.}
{\bf 48}, 119-130 (1976).

\bibitem[S7]{Hu2016}  Y. Hu, P. Zoller, and J. C. Budich, Dynamical buildup of a quantized hall response from
nontopological states. {\em Phys. Rev. Lett.} {\bf 117}, 126803 (2016).

\end{thebibliography}
\end{document}